\documentclass[10pt,a4paper]{article}

\usepackage{arxiv}

\usepackage[round,sort&compress,numbers]{natbib}

\usepackage[utf8]{inputenc}
\usepackage[T1]{fontenc}
\usepackage{graphicx}
\usepackage{lipsum}
\usepackage[table, dvipsnames]{xcolor}
\usepackage{amsfonts}
\usepackage{bm}
\usepackage{amsmath}
\usepackage{amsthm}
\usepackage{mathtools, nccmath}
\usepackage{url}
\usepackage{booktabs}
\usepackage{tabularx}
\usepackage{nicefrac}
\usepackage{microtype}
\usepackage[hidelinks]{hyperref}
\usepackage[inline]{enumitem}
\usepackage{float}
\usepackage{siunitx}
\usepackage{wrapfig}
\usepackage[font=footnotesize, labelfont=bf]{caption}
\usepackage{subcaption}
\usepackage{adjustbox}
\usepackage{xparse}
\usepackage{tikz}
\usetikzlibrary{shapes.geometric}
\usepackage{changepage}
\usepackage{afterpage}
\usepackage{etoc}
\usepackage{placeins}
\usepackage{mfirstuc}
\usepackage{threeparttable}
\usepackage{makecell}
\usepackage{ragged2e}
\usepackage[framemethod=TikZ]{mdframed}
\usepackage[symbol]{footmisc} %
\usepackage{multicol}
\usepackage{xspace}
\usepackage{multirow}
\usepackage[group-separator={,}]{siunitx} %
\allowdisplaybreaks %

\usepackage{Definitions}
\definecolor{lightgrey}{rgb}{0.43,0.43,0.43}
\hypersetup{
    colorlinks,
    anchorcolor={blue!50!black},
    linkcolor={blue!50!black},
    citecolor={blue!50!black},
    urlcolor={black}
}

\usepackage{Definitions}
\newcommand\ours{LCD\xspace}
\newcommand\ourss{LCDs\xspace}
\newcommand\oursbase{latent causal diffusion}
\newcommand\Oursbase{Latent causal diffusion}

\newcommand\oursfull{\oursbase\xspace}

\newcommand\Oursfulls{\Oursbase s\xspace}

\newcommand\oursanalysis{CLIPR\xspace}

\newcommand\ourscombo{\ours-\oursanalysis}

\newcommand{\C}{{C}\xspace}
\newcommand{\DE}{{DE}\xspace}
\newcommand{\DEs}{{DEs}\xspace}

\newcommand{\subp}[1]{(\textbf{#1})\xspace}
\newcommand{\subpp}[1]{\textbf{#1}\xspace}

\newcommand\scrna{scRNA-seq\xspace}
\newcommand\perturbseq{Perturb-seq\xspace}

\newcommand\methodsname{Materials and Methods\xspace}
\newcommand{\methods}{\hyperref[sec:methods]{\methodsname}\xspace}
\newcommand{\secref}[2]{\hyperref[#1]{#2}\xspace}

\newcommand\refappname{SI Appendix}
\newcommand{\refapp}{\hyperref[sec:appendix]{\refappname}\xspace}

\newcommand{\refappfig}[1]{\hyperref[#1]{\refappname, Fig.~\ref*{#1}}}
\newcommand{\refappfigs}[2]{\hyperref[#1]{\refappname, Figs.~\ref*{#1} \& \ref{#2}}}
\newcommand{\refapptab}[1]{\hyperref[#1]{\refappname, Table~\ref*{#1}}}

\NewDocumentCommand{\figref}{m o}{%
  (Fig.~\hyperref[#1]{\ref*{#1}\IfValueT{#2}{#2}})%
}
\NewDocumentCommand{\ffigref}{m o}{%
  Fig.~\hyperref[#1]{\ref*{#1}\IfValueT{#2}{#2}}%
}

\renewcommand{\eqref}[1]{\textup{(Eq.~\ref{#1})}}

\newcommand\erdosrenyi{\text{Erd{\H{o}}s-R{\'e}nyi}\xspace}
\newcommand{\dcor}{\mathrm{dcorr}\xspace}

\newcommand{\gimag}{\mathrm{GI}_\mathrm{mag}}
\newcommand{\gisim}{\mathrm{GI}_\mathrm{sim}}
\newcommand{\giequality}{\mathrm{GI}_\mathrm{eq}}
\newcommand{\gifit}{\mathrm{GI}_\mathrm{fit}}

\newcommand{\zip}{\mathrm{ZIP}}
\newcommand\ratefunction{\mu}
\newcommand{\logp}{\mathrm{log}\num{1}\mathrm{p}}
\newcommand{\unif}{\mathrm{Unif}} 
\newcommand{\unifpm}{\mathrm{Unif}_{\pm}}

\newcommand{\dt}{\mathrm{d}t}
\newcommand{\dWt}{\mathrm{d}\mathbb{W}(t)}
\newcommand{\dd}{\mathrm{d}}
\newcommand{\bzero}{\boldsymbol{0}}
\newcommand{\deltab}{\boldsymbol{\delta}}

\newcommand{\xbc}{\widetilde{\xb}\xspace}
\newcommand{\dimdrift}{d_f}

\newcommand{\pert}{q}
\newcommand{\pertemb}{\eb_{\pert}}
\newcommand{\pertembprime}{\eb_{\pert'}}

\newcommand{\flin}{\widetilde{f}}
\newcommand{\fzero}{\vb}
\newcommand{\Fzero}{\Vb}
\newcommand{\finf}{\lb}
\newcommand{\Finf}{\Lb}

\newcommand{\flinpert}{\flin_\pert}

\newcommand{\ffzero}[1]{\fzero_{#1}\xspace}
\newcommand{\ffinf}[1]{\finf_{#1}\xspace}

\newcommand{\fzeropert}{\ffzero{f_\pert}}

\newcommand{\fzeropertstart}{\ffzero{f_{\pert_1}}}
\newcommand{\fzeroperti}{\ffzero{f_{\pert_i}}}
\newcommand{\fzeropertend}{\ffzero{f_{\pert_k}}}

\newcommand{\finff}{\ffinf{f}}
\newcommand{\finfpert}{\ffinf{f_\pert}}

\newcommand{\finfpertstart}{\ffinf{f_{\pert_1}}}
\newcommand{\finfperti}{\ffinf{f_{\pert_i}}}
\newcommand{\finfpertend}{\ffinf{f_{\pert_k}}}

\newcommand{\Ahat}{\smash{\widehat{\Ab}}}

\newcommand{\scalingmat}{\Gammab}

\newcommand{\transform}{\phi}
\newcommand{\tbias}{\mb}
\newcommand{\tmat}{\Tb}

\newcommand{\dsmscale}{\alpha}
\newcommand{\Dsmscale}{\bm{\alpha}}

\newcommand{\Endparasplit}{}
\renewcommand\methodsname{Methods\xspace}

\usepackage{titling}
\pretitle{\vspace{-50pt}\begin{center}\Large\bfseries}
\posttitle{\end{center}\vspace{10pt}}
\preauthor{\begin{center}\large}
\postauthor{\end{center}\vspace{-20pt}}
\title{Latent Causal Diffusions for Single-Cell Perturbation Modeling}

\author[a]{Lars Lorch}
\author[b,c]{Jiaqi Zhang}
\author[d,e]{Charlotte Bunne}
\author[a,*]{\\Andreas Krause}
\author[f,g,a,*]{Bernhard Sch{\"o}lkopf}
\author[b,c,*]{Caroline Uhler}

\affil[a]{Department of Computer Science, ETH Z{\"u}rich, Z{\"u}rich, Switzerland}
\affil[b]{Laboratory for Information and Decision Systems, Massachusetts Institute of Technology, Cambridge, MA, USA}
\affil[c]{Eric and Wendy Schmidt Center, Broad Institute of MIT and Harvard, Cambridge, MA, USA}
\affil[d]{School of Computer and Communication Sciences, EPFL, Lausanne, Switzerland}
\affil[e]{Swiss Institute for Experimental Cancer Research, School of Life Sciences, EPFL, Lausanne, Switzerland}
\affil[f]{Max Planck Institute for Intelligent Systems, T{\"u}bingen, Germany}
\affil[g]{ELLIS Institute, T{\"u}bingen, Germany\vspace{3pt}}

\affil[*]{Correspondence to: krausea@ethz.ch, bs@tuebingen.mpg.de, cuhler@mit.edu}

\begin{document}

\maketitle

\begin{abstract}

Perturbation screens hold the potential to systematically map regulatory processes at single-cell resolution, yet modeling and predicting transcriptome-wide responses to perturbations remains a major computational challenge.
Existing methods often underperform simple baselines,
fail to disentangle measurement noise from biological signal,
and provide limited insight into the causal structure governing cellular responses.
Here, we present the latent causal diffusion (LCD), a generative model that frames single-cell gene expression as a stationary diffusion process observed under measurement noise.
LCD outperforms established approaches in predicting the distributional shifts of unseen perturbation combinations in single-cell RNA-sequencing screens
while simultaneously learning a mechanistic dynamical system of gene regulation.
To interpret these learned dynamics,
we develop an approach we call causal linearization via perturbation responses (CLIPR),
which yields an approximation of the direct causal effects between all genes modeled by the diffusion.
CLIPR provably identifies causal effects under a linear drift assumption and recovers causal structure in both simulated systems and a genome-wide perturbation screen, where it clusters genes into coherent functional modules and resolves causal relationships that standard differential expression analysis cannot.
The LCD-CLIPR framework bridges generative modeling with causal inference to predict unseen perturbation effects and map the underlying regulatory mechanisms of the transcriptome.

\end{abstract}

\section*{Introduction}

Genome-wide perturbation screens now enable causal interrogation of gene function at single-cell  resolution \citep{replogle2022mapping,carlson2023genome}.
As these screens scale to capture the transcriptomic effects of thousands of individual gene perturbations, the central challenge shifts from data collection to modeling regulatory interactions and predicting the outcomes of multi-gene combinations.
Genetic interactions are often nonlinear, with the effects of combined perturbations deviating from the sum of their individual effects \citep{rood2024toward,norman2019exploring}.
Because the combinatorial search space grows exponentially, exhaustive experimental assays quickly become infeasible \citep{wessels2023efficient,cleary2020necessity}.
Mapping the genetic circuitry of cellular programs thus requires computational causal models that can generalize from observed data to predict the effects of unseen perturbation combinations and reveal the mechanisms governing their interactions
 \citep{rood2024toward,tejada2025causal,uhler2024building,taylor2024future,dimitrov2026interpretation}.

While deep learning has advanced representation learning for single-cell biology, current models often struggle to outperform simple, domain-aware baselines when predicting post-perturbation responses 
\citep{gavriilidis2024mini,ahlmann2025deep,gaudelet2024season}.
Foundation models pre-trained on large single-cell atlases, for instance, frequently fail to predict gene expression changes for held-out perturbations more accurately than additive heuristics \citep{theodoris2023transfer,cui2024scgpt,rosen2024toward}.
Neural network models trained on individual perturbation screens typically focus on average effects, neglecting the heterogeneity and stochasticity that define cellular states \citep{cui2024scgpt,roohani2024predicting}.
Furthermore, these models rarely account for measurement noise, often conflating technical artifacts with biological signal \citep{bunne2023learning,lotfollahi2023predicting,he2025morph}.
Above all, these architectures remain largely opaque, failing to characterize how perturbations propagate through genetic pathways to drive transcriptional changes.
Consequently, current methods provide limited structural insight into the causal mechanisms that govern cellular responses.

\Endparasplit %

In this work, we introduce single-cell perturbation modeling based on the \oursfull (\ours), 
a generative model that frames gene expression as a latent stationary dynamical system observed under measurement noise \citep{weinreb2018fundamental,teschendorff2021statistical}.
We formalize gene regulation as a diffusion process governed by a system of stochastic differential equations (SDEs), where each gene's expression evolves mechanistically as a function of its regulators.
By assuming this diffusion process operates in equilibrium \citep{gillespie1977exact,kepler2001stochasticity,raj2006stochastic,shahrezaei2008analytical}, 
we can infer the functional causal mechanisms of the SDEs without direct observation of the temporal dynamics,
viewing single-cell data as samples from the stationary distribution under a measurement model \citep{lopez2018deep,dibaeinia2020sergio,lorch2024causal}.
Unlike previous deep learning approaches, this enables identifying causal data-generating mechanisms while explicitly decoupling biological stochasticity from technical noise.
We model perturbations of the transcriptome as modifications to the underlying gene regulatory mechanisms, causing the diffusion to reach a different equilibrium.
When applied to large-scale \perturbseq screens \citep{wessels2023efficient,norman2019exploring}, \ourss predicted the distributional shifts of unseen perturbation combinations more accurately than established approaches while learning an explicit, gene-level dynamical system as the generative model.

The regulatory dynamics learned by \ourss are parameterized by neural networks, so we render their causal dependencies interpretable by developing \oursanalysis, a linearization technique that approximates the direct gene-gene causal effects modeled in the diffusion process.
For linear systems, \oursanalysis provably identifies causal effects by leveraging how drift perturbations behave at a control state and in equilibrium.
In nonlinear systems, it provides the minimum-norm linear causal effects that best approximate the  drift's response to perturbations.
We validated \oursanalysis on simulated systems, demonstrating that it reliably recovers ground-truth causal structures from complex neural network drifts.  
Finally, by applying \ours and \oursanalysis to a genome-wide \perturbseq assay \citep{replogle2022mapping}, we uncovered regulatory mechanisms in K562 cells that remain hidden under standard differential expression analysis, providing a scalable framework for mapping causal dependencies in the cell.

\begin{figure*}[t]
    \centering
    \includegraphics[
        width=\textwidth,
        trim={1pt 165pt 3pt 3pt}, clip
    ]{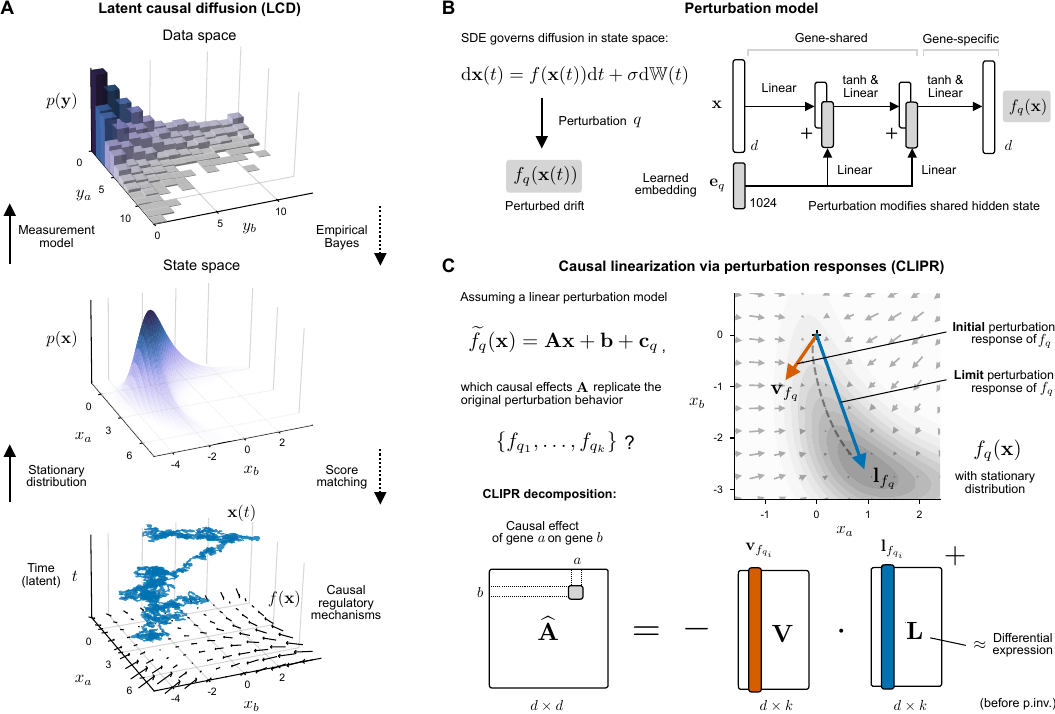}
    \caption{%
    \textbf{%
    The \oursfull (\ours) model and \oursanalysis.} 
    \subp{A} \ourss model single-cell expression data $\yb$ (top) as noisy observations of unobserved, noiseless gene states $\xb$ (center).
    States $\xb$ are samples from a stationary stochastic process $\xb(t)$ evolving under the causal regulatory dynamics $f(\xb)$ of an SDE (bottom).
    To train \ourss, we first infer $p(\xb)$ via empirical Bayes, then learn $f$ in state space via score matching. 
    \subp{B} Dynamics $f$ explicitly model gene regulation and are parameterized by a neural network with shared hidden state across genes.
    Perturbations modify this hidden state via learned embeddings $\pertemb$. 
    \subp{C} To interpret the causal dependencies in $f$, 
    we estimate a linear causal matrix that induces the perturbation behavior of $f$ (left).
    This matrix is computed from the initial ($\fzeropert$) and limit ($\finfpert$) responses when $f$ is perturbed.
    The limit response is approximated by following the vector field flow until convergence (right).
    }
    \label{fig:method}
\end{figure*}

\section*{Results}\label{sec:results}

\subsection*{A causal generative model of the transcriptome}

\Oursfulls model the joint expression of $d$ genes in a cell 
as a stochastic process $\xb(t)\in \RR^d$ evolving over time $t$, as gene expression is inherently stochastic \citep{kepler2001stochasticity,elowitz2002stochastic,ozbudak2002regulation}.
Each gene $g$ is represented by a random state $x_g(t)$, corresponding to a noiseless, unobserved version of expression, that evolves according to 
a stochastic differential equation (SDE) \citep{oksendal2003stochastic}
\begin{align}\label{eq:causal-model}
    \dd x_g(t) &=  \big[f(\xb(t))\big]_g  \dd t + \sigma \dd\WW_g(t) \, ,
\end{align}
where $\WW_g(t)$ denotes standard Brownian motion.
The drift function $f : \RR^{d} \rightarrow \RR^{d}$ models the causal regulatory mechanisms among the genes, and the diffusion term captures stochastic fluctuations with scale $\sigma$.
The joint stochastic process $\xb(t)$, obtained by solving the coupled SDEs for all genes, defines the distribution over cell states at time $t$ \figref{fig:method}[A, bottom].

\ourss model the process $\xb(t)$ as {\em stationary}, so the SDEs induce a time-independent  distribution $p(\xb)$ over gene states. 
Steady-state models are canonically used as mathematical descriptions of gene expression
\citep{gillespie1977exact,kepler2001stochasticity,raj2006stochastic,shahrezaei2008analytical}.
The stationarity assumption enables inference of the causal dynamics $f$ directly from the state distribution $p(\xb)$, without requiring time-resolved observations of $\xb(t)$ itself \figref{fig:method}[A, center] \citep{weinreb2018fundamental,lorch2024causal}.
The time dimension of \ourss allows modeling gene feedback loops
by implicitly unrolling causal effects over time, 
akin to biological systems
\citep{thomas1995dynamical,elowitz2000synthetic,alon2007network}.
This concept of time differs from recent diffusion-based models,
which either model time series  \cite{yeo2021generative,vinyard2023scdiffeq}
or treat time as computation steps rather than a physical timescale of gene regulation \citep{he2025squidiff}.

We parameterize the drift $f$ by a neural network that shares parameters across the causal mechanisms of each gene (\ffigref{fig:method}[B]).
Specifically, $[f(\xb)]_{g}$ and $[f(\xb)]_{g'}$ share a hidden state, reflecting the substantial structure of gene regulatory circuits
\citep{tanay2002discovering,segal2003module,isalan2008evolvability,heimberg2016low}.
In our experiments, the shared hidden dynamics enabled \ourss to scale to modeling a thousand genes jointly.
We model perturbations as shifts in this hidden state, computed from learned embedding vectors $\pertemb$, where $\pert$ is the index of a perturbation \figref{fig:method}[B],
as gene expressions tend to co-vary across perturbations, and perturbations cluster by their shared transcriptomic effects
\citep{tanay2002discovering,segal2003module,hughes2000functional}.
Combinations of perturbations are represented as additive compositions of these embeddings. 
Perturbations thus act on the regulatory mechanisms of $f$ rather than directly on gene expression, forcing the diffusion to converge to a new steady state.
We write $f_{\pert}$ to denote the dynamics $f$ under perturbation $\pert$.

We do not observe the state distribution $p(\xb)$ directly but instead through noisy measurements of single cells.
\ourss model each single-cell count vector $\yb \in \NN^d$ as an independent sample from the latent stationary diffusion, observed under a zero-inflated Poisson ($\zip$) likelihood \citep{dibaeinia2020sergio,lorch2024causal}
\begin{align}\label{eq:noise-model}
    p(y_g \given x_g; \pi_g) &= \zip (y_g;  \ratefunction_g(x_g), \pi_g) %
\end{align}
with dropout probability $\pi_g$.
Here, $\ratefunction_g: \RR \rightarrow \RR^+$ is a link function with fixed, gene-specific scaling that maps states $x_g$ to nonnegative Poisson rates.
If the Poisson rate follows a Gamma distribution, the distribution of $y_g$ has a zero-inflated negative binomial form,
which corresponds to the steady-state of the promoter-activation model commonly characterizing \scrna data \citep{shahrezaei2008analytical,grun2014validation,lopez2018deep}.
\ourss generalize this to richer distributions $p(\yb)$ by capturing causal dependencies among genes in $p(\xb)$ \figref{fig:method}[A, top].

To infer the model from data, we view a perturbation screen as a collection of count distributions, one per perturbation $\pert_i$, where $\pert_0$ represents  the unperturbed setting ($f_{\pert_0} = f$).
Each count distribution arises from the same drift and measurement model, differing only by the perturbation-specific shifts in the hidden state of $f_{\pert_i}$ \citep{lorch2024causal}.
To separate measurement and biological noise, we first infer the likelihood parameters $\pib$ and state distribution $p(\xb)$ for each perturbation condition.
We then jointly fit the dynamics $f$ and perturbation embeddings $\pertemb$ to all inferred state distributions (\methods).

\subsection*{Estimating causal dependencies in stationary diffusions}

A learned \ours models a causal generative process, in which genes mechanistically regulate one another, but the dynamics of \ourss are not directly interpretable, since the drift $f(\xb(t))$ is nonlinear and depends on $\xb(t)$.
However, perturbations of the drift $\{f_{\pert_1}, \dots, f_{\pert_k}\}$ reveal key regulatory mechanisms modeled by $f$.
We can leverage this perturbation behavior to obtain a linearization of the system that preserves its overall perturbation responses, enabling us to interpret the causal structure of a learned \ours drift $f$.

The key idea is to ask: what {\em linear} drift  $\flin$ would replicate the perturbation behavior of $\{f_{\pert_1}, \dots, f_{\pert_k}\}$ \figref{fig:method}[C, left]?
We show that such a linear approximation can be estimated in closed form from simple properties of the perturbed drifts $f_{\pert_i}$.
Since linear systems describe gene-gene regulation by a single parameter per pair, they yield an interpretable approximation of the regulatory dynamics of $f$.
We consider linear SDEs \eqref{eq:causal-model} with drift
\begin{align}\label{eq:linear-sde}
    \flinpert(\xb)&= \Ab\xb + \bb + \cb_\pert \, ,
\end{align}
where $\Ab \in \RR^{d \times d}$ and $\bb \in \RR^{d}$ are causal effect and bias parameters,
$\cb_\pert \in \RR^{d}$ models a perturbation $q$,
and the real part of the maximum eigenvalue $\Re(\lambda_{\max}(\Ab))$ is strictly negative.
Linear SDEs of this form have a unique Gaussian stationary distribution with mean \citep{jacobsen1993brief}
\begin{align}\label{eq:linear-sde-mean}
    \mub_\pert = - \Ab^{-1}(\bb + \cb_\pert) \, .
\end{align}
We can describe the perturbation $f_\pert$ of a drift $f$ by two intuitive properties: 
the drift vector acting at a control state and the steady state it induces.
To formalize these properties, we use the zero vector $\bzero$ as the control state, which represents the mean of a standardized, unperturbed linear system.
We then define the {\em initial perturbation response} of $f_{\pert}$ as
\begin{align}\label{eq:fzero-vector}
	\fzeropert := f_{\pert}(\bzero) 
	\, ,
\end{align}
and the {\em limit perturbation response} of $f_{\pert}$ as the fixed point
\begin{align}\label{eq:finf-vector}
	\finfpert 
	&:= \xb^* 
	\quad\text{s.t.}\quad
	f_{\pert}(\xb^*) = \bzero 
	\, .
\end{align}
For the linear model, the post-perturbation mean $\mub_\pert$ is the fixed point $\smash{\finfpert}$.
Therefore, the limit perturbation response $\smash{\finfpert}$ directly corresponds to differential expression, assuming the unperturbed system has mean $\bzero$.
By contrast, the initial perturbation response $\smash{\fzeropert}$ represents the initial drift towards the post-perturbation steady state \figref{fig:method}[C, right].
We assume that a fixed point exists and can be approximated by integrating the drift flow (\methods).

Applying different perturbations 
 $\pert_i$ 
to a diffusion drift $f$
yields different perturbation responses 
 $\fzeroperti$ and $\finfperti$. 
Each respective perturbation outcome reveals key drivers of the dynamics $f$, and collectively they map how the system behaves under perturbations.
Our theorem below shows that,
given several perturbation response vectors,
we can construct a linear causal effect matrix $\Ahat$ that replicates the same response vectors as the original perturbed drifts.
Moreover, assuming the true drift is linear, sufficiently diverse perturbations identify all causal effects:
\begin{theorem}\label{theorem:clipr}
Let $\smash{\fzeroperti}$ and $\smash{\finfperti}$ be  perturbation responses of a general drift $\smash{f_\pert}$ for perturbations $\pert_i \in \{\pert_1, \dots, \pert_k\}$.
The linear drift matrix (Eq.~\ref{eq:linear-sde}) 
that approximates these perturbation response vectors 
with least-squares error 
and minimum Frobenius norm
is given by
\begin{align}\label{eq:clipr-estimator}
    \Ahat = - \Fzero \Finf^+  \, ,
\end{align}
where $\smash{(\cdot)^+}$ denotes the Moore-Penrose pseudoinverse, and $\smash{\Fzero, \Finf \in \RR^{d \times k}}$ 
\vspace*{-0.8pt}are matrices with $i$-th columns ${\fzeroperti}$ and ${\finfperti}$, respectively.
\vspace*{-1pt}Moreover, if the drift function is linear (${f_\pert = \flin_\pert}$) and $\Finf$ has full row rank, then $\Ahat = \Ab$.
\end{theorem}
Our proof is given in \refapp.
When the bias $\bb$ is generic, a sufficient condition for $\Finf$ being full rank is that $k \geq d$ perturbation shift vectors $\cb_\pert$ are linearly independent.
We refer to this system as the causal linearization via perturbation responses (\oursanalysis).
In practice, we use a Tikhonov-regularized variant,
which improves stability under model mismatch by trading off perturbation response fit against smaller norms of the causal effects $\Ab$ via a scalar $\lambda$ (see proof of Theorem \ref{theorem:clipr}):
\begin{align}\label{eq:clipr-estimator-regularized}
	\Ahat_\lambda =
    - \Fzero 
    \Finf^\top 
    \big (
    	\Finf\Finf^\top 
    	+ \lambda \Ib
    \big )^{-1}
    \, . 
\end{align}

Theorem \ref{theorem:clipr} has several implications.
First, 
the identification result suggests that all single-gene perturbations can fully disentangle the causal effects in a linear system.
Specifically, Theorem \ref{theorem:clipr} shows that linear causal effects can be directly decomposed into (and uniquely identified by)  the \DEs~$\Finf$ and the drift forces $\Fzero$ at the control state $\bzero$ \figref{fig:method}[C, right].
While access to exact perturbation responses constitutes an idealized setting, the result provides a theoretical foundation for  estimating $\Ab$ by computing 
$\fzeropert$ and $\finfpert$ for drifts $f_\pert$ learned from finite samples and noisy observations.

Second, 
the simple form $\Ahat = - \Fzero \Finf^+$ highlights a fundamental distinction between direct causal effects and differential expression.
Under the simplifying assumptions discussed above, differential expression (\DE), the mean change in expression induced by a perturbation, corresponds to the limit perturbation responses $\Finf$.
\oursanalysis shows that \DE, as captured by $\Finf$, is distinct from the direct causal effects among the genes themselves: \DE emerges from unrolling direct causal effects $\Ab$ over time. 
We say that gene $a$ is a direct cause of gene $b$ if $b$ depends on perturbations of $a$ when all other genes are held fixed \citep{mooij2016distinguishing}.
By definition of the drift \eqref{eq:linear-sde}, $\Ab$ encodes precisely these direct causal dependencies.

Third, \oursanalysis enables interpreting nonlinear drifts $f_\pert$, since $\fzeropert$ and $\finfpert$ can be computed for any drift function that models perturbations.
\oursanalysis then reveals the linear causal effects $\Ahat$ that best approximate the perturbation behavior of the original dynamics $f_\pert$.
The regularized estimator $\Ahat_\lambda$ mitigates artifacts resulting from model mismatch when the linear model is overdetermined by the nonlinear responses, since it encourages smaller norms of $\Ab$ in favor of overfitting the responses.

Fourth, the perturbation shifts $\cb_\pert$ in the approximated linear drift \eqref{eq:linear-sde} are {\em not} needed to compute the estimate $\Ahat$. 
What matters is that the original drift $f_\pert(\xb)$ can be evaluated and integrated to obtain $\Fzero$ and $\Finf$.
This also implies that perturbations in the approximated linear drift \eqref{eq:linear-sde} neither need to have known targets nor affect single genes and could therefore also be, for example, induced by drugs.
In fact, the following corollary shows that the perturbation shifts $\cb_\pert$ in the approximated linear model are implied by the \oursanalysis matrix $\Ahat$:
\begin{corollary}\label{corollary:clipr-vectors}
The bias $\widehat{\bb}$ and shift vectors $\widehat{\cb}_\pert$ of the linear model (Eq.~\ref{eq:linear-sde}) implied by the matrix $\Ahat$ 
of Theorem \ref{theorem:clipr}
are 
\begin{align*}
    \widehat{\bb} &= -\Ahat \finff
    & 
    \text{and}
    &&
    \widehat{\cb}_\pert &= \Ahat \finff  -  \Ahat \finfpert \, .
\end{align*}
\end{corollary}

\begin{figure*}[ht!]
    \centering
    \includegraphics[
        width=\textwidth,
        trim={1pt 168pt 0pt 2pt}, clip
    ]{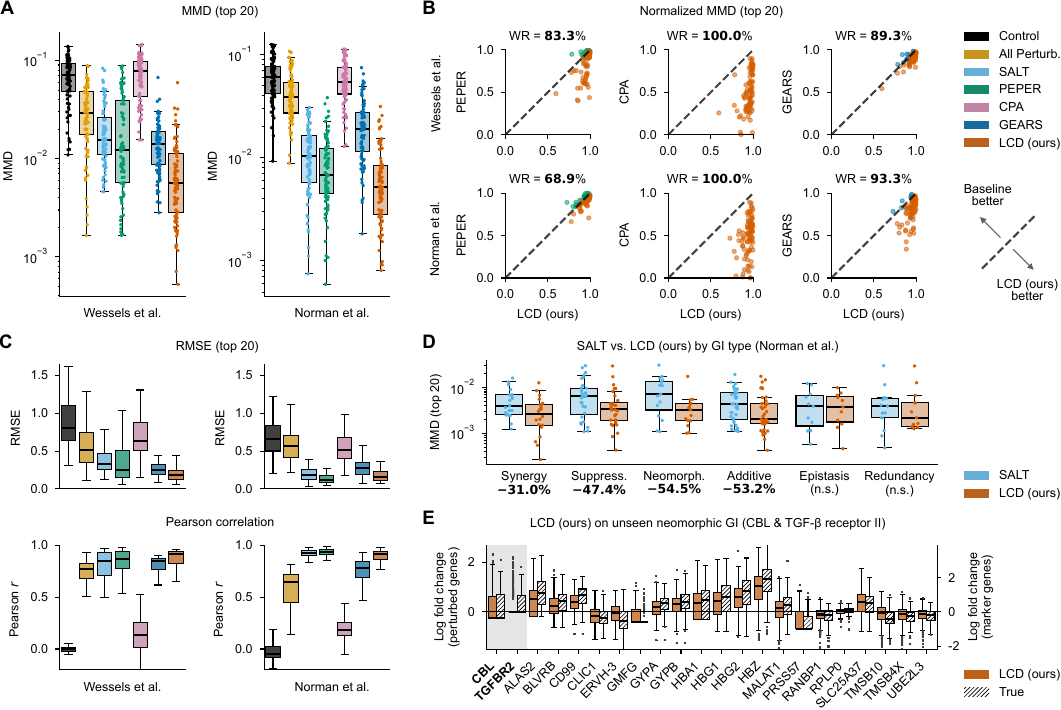}
    \caption{%
    \textbf{\ourss outperform state-of-the-art models at predicting combinatorial perturbation effects.} 
    \subp{A}~Maximum mean discrepancy (MMD) between predicted and observed single-cell expressions for held-out double-gene perturbations of the two datasets, computed on top 20 differentially expressed (\DE) genes. 
    Each dot represents one perturbation.
    Two leftmost bars show baseline performance of randomly sampled control and perturbed cells from all conditions.
    \subp{B} 
    Per-perturbation comparison of \ours with deep-learning baselines on MMD, normalized to $[0,1]$ across methods.
    Win rate (WR): proportion of perturbations where \ours achieved lower error.
    \subp{C} 
    Root mean squared error (RMSE) of predicted mean expression of top 20 \DE genes (top) and Pearson correlation on full mean expression vector (bottom)
    \subp{D} 
    MMD stratified by genetic interaction (GI) type, comparing \ours and the additive baseline SALT. 
    Axis labels indicate significant improvement per GI category (percentage reduction in median MMD by \ours) or not (n.s.; one-sided Wilcoxon signed-rank test, $P < 0.05$). 
    \subp{E} 
    Predicted vs.\ observed expression for top 20 \DE genes of a held-out perturbation with neomorphic  GI. 
    Grey section shows the perturbed genes (bold labels).
    All box plots show the median, interquartile range (IQR), and whiskers extend to farthest points within $1.5 \times$ IQR.
    }
    \label{fig:benchmarking}
\end{figure*}

\subsection*{Perturbation effect prediction at single-cell resolution}

We evaluated \ourss against established methods for predicting the effects of unseen perturbation combinations in \perturbseq assays.
Our evaluation used two datasets that contain both single-gene and double-gene perturbations,
spanning different cell lines and CRISPR-Cas technologies 
(Norman et al.\ \citep{norman2019exploring} and Wessels et al.\ \citep{wessels2023efficient}; \refapptab{tab:datasets}).
For both datasets, each method was trained on control samples, all single-gene perturbations, and a subset of double-gene perturbations.
After training, methods predicted full single-cell distributions of the transcriptome for the held-out two-gene combinations. 
Using ten disjoint testing folds, we obtained held-out predictions for all two-gene perturbations in both screens (\refappfig{fig:illustration-benchmark-splits}).

We compared \ourss to various methods.
Recent findings suggested that established approaches perform on par with additive heuristics \citep{gavriilidis2024mini,ahlmann2025deep,li2024systematic}.
We thus evaluated an additive baseline called SALT and its variant PEPER, which learns a nonlinear correction on top of SALT \citep{gaudelet2024season}.
We also evaluated the deep-learning approaches
CPA \citep{lotfollahi2023predicting}, which uses autoencoders,
and GEARS \citep{roohani2024predicting},
which learns graph neural networks over gene relationship graphs sourced from prior knowledge
(\methods).
We quantified prediction accuracy relative to the observed post-perturbation data using several metrics.
Our primary metric was the maximum mean discrepancy (MMD), which measures full distributional fit \citep{gretton2012kernel,bunne2023learning}.
Low MMD implies the prediction matches all moments of the observed single-cell distribution, characterizing not only its means, but also its (co)variances, skewness, and tail-heaviness.
Following prior work, we also report the root mean squared error (RMSE) and Pearson correlation of the means \citep{lotfollahi2023predicting,roohani2024predicting}.
We mostly focused on metrics computed on the top \num{20} differentially expressed genes, as most genes often show negligible change under perturbation \citep{bunne2023learning,roohani2024predicting}.

\ourss significantly outperformed both the heuristics and established approaches in terms of overall distributional accuracy of the predicted single-cell transcriptomes  \figref{fig:benchmarking}[A].
Per-perturbation comparisons on each dataset show that \ourss made more accurate predictions as measured by MMD than all other methods in more than roughly 80\% of test perturbations \figref{fig:benchmarking}[B].
Only PEPER achieved comparable performance to \ourss on the screen by Norman et al., where more single-gene perturbations are available for calibrating its additive heuristic (98 for Norman et al.\ versus 5 for Wessels et al.; \refapptab{tab:datasets}),
suggesting that \ourss disentangle individual gene contributions effectively from combined perturbations.
When reducing the single-cell transcriptome to a mean expression vector,
\ourss still provided the most accurate predictions across all baselines, though SALT and PEPER performed more competitively \figref{fig:benchmarking}[C].
\ourss thus achieved significant accuracy gains by modeling a full singe-cell count distribution.

Combinations of gene perturbations may behave differently than their individual components suggest.
We followed Norman et al.\ \citep{norman2019exploring} and classified double-gene perturbations into six genetic interaction (GI) types based on their observed transcriptomic profiles, stratifying the performance of \ourss by GI type (\refapp).
By comparing to the na\"{\i}ve additive SALT heuristic,
we isolated the GI types for which \ours yielded significant improvement \figref{fig:benchmarking}[D].
\ourss improved most over an additive model, for example, for unexpected (neomorphic) GIs, where a linear model of the individual effects does not provide a good fit for the combination profile.
By contrast, SALT performed on par with \ourss for redundant or epistatic GIs, where either both or one of the individual perturbations, respectively, strongly explain the combined effect.
While both methods use additivity to model combinations, they differ in where this addition occurs: 
SALT na\"{\i}vely sums differential expression profiles, whereas \ours sums perturbation embeddings $\pertemb$ that modulate the hidden state of a nonlinear drift function (\ffigref{fig:method}[A and B]). 
This allows \ourss to capture interaction effects that unroll through the nonlinear dynamics.
\ffigref{fig:benchmarking}[E] illustrates the \ours prediction on a two-gene perturbation combination  with established GI that was not seen during training 
\citep{zuo2013cbl}.
\ours correctly predicted most observed \DE trends, both of the on-target and marker (top 20 DE) genes.
\refappfig{fig:de-gi-supplementary} provides additional plots for the other GI types.

\begin{figure*}[t]
    \centering
    \includegraphics[
        width=\textwidth,
        trim={1pt 262pt 3pt 1pt}, clip
    ]{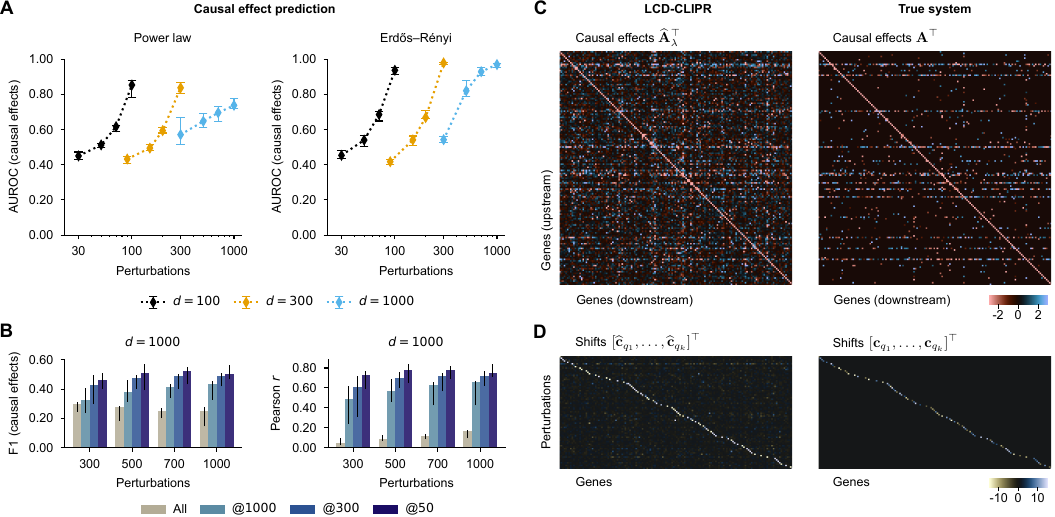}
    \caption{%
    \textbf{\ourscombo accurately recovers causal effects in linear systems from learned diffusion drifts.} 
    \subp{A}~AUROC for classifying gene-gene causal effects (positive, negative, or absent) across all gene pairs.
    Results are shown for systems of 100, 300, and 1000 genes across varying training perturbation counts ($x$-axis), comparing power-law (left) and \erdosrenyi (right) regulatory dependencies in $\mathbf{A}$. 
    \subp{B} F1 score for classifying gene-gene causal effects (left) and average Pearson correlation between predicted and true causal effects (right) for the top $k$ predicted effects, ranked by absolute magnitude, under power-law dependencies. 
	Markers indicate medians, error bars indicate 10--90th percentiles across test systems. 
    \subp{C} Comparison of causal effects inferred by \ourscombo vs.\ the ground-truth data-generating system (300 genes).
    \subp{D} Inferred and true perturbation vectors for the system shown in \textbf{C}.
    For visual clarity, only 80 perturbations and 150 genes with the largest outgoing effects are displayed.
    }
    \label{fig:linear}
\end{figure*}

\subsection*{Recovering causal effects in simulated systems}

Next, we evaluated the \oursanalysis estimator component of our model.
A quantitative evaluation of causal effect predictions is challenging, because the ground-truth links in gene networks are not generally known.
To enable an accurate validation, we first tested \oursanalysis on perturbation data simulated by linear systems with single-gene shift perturbations \eqref{eq:linear-sde}, where the true causal structure and perturbation effects are known (\methods).
Here, we omitted the measurement model component of \ourss \eqref{eq:noise-model} and directly trained the neural network drift $f$ on samples of the linear systems to validate the causal structure component in isolation.
Our evaluation covered ground-truth systems with varying numbers of genes and perturbed genes as well as regulatory structures in $\Ab$, testing ten randomly-generated systems for each setting.
We measured accuracy by classifying the gene-gene causal effects $\Ahat_\lambda$ predicted by \oursanalysis into positive, negative, or absent and then computing the area under a (three-class) receiver operating characteristic curve (AUROC) given the ground-truth $\Ab$ (\methods).
We also tested how the predicted effect strength of $\Ahat_\lambda$ influences accuracy by computing Pearson correlations of predicted and true effects as well as F1 classification scores for the top $k$ (@$k$) predicted effects by absolute value.
We use \ourscombo to denote the linear causal effects estimated by applying \oursanalysis to the learned \ours drifts.

\ourscombo achieved high accuracy in predicting both causal links and their effect signs across all sparsity structures, numbers of genes, and numbers of perturbations \figref{fig:linear}[A].
The classification accuracy of \ourscombo effects monotonically improved
with the number of perturbations,
reaching median \mbox{AUROC} values between 
0.74--0.84 (power-law structures) 
and 0.93--0.98 (\erdosrenyi structures)
when observing $d$ perturbations. 
Moreover, stronger predicted effects resulted in higher F1 scores and correlations with the ground truth \figref{fig:linear}[B], indicating that the strongest perturbation effects learned by the \ours drift $f_\pert$ aligned strongly with those of the true system. 
We later build on this finding when applying \ourscombo to a genome-wide \perturbseq screen.

\ffigref{fig:linear}[C] depicts 
the \oursanalysis effects extracted from a \ours drift 
alongside those of the true data-generating system.
\ourscombo correctly identified key regulator genes of the power-law regulatory structure, visible as horizontal non-zero lines with many effects.
The stability-inducing negative diagonal is recovered accurately, but \ourscombo tends to be less sparse, since the estimator minimizes an $\ell^2$ norm (Theorem \ref{theorem:clipr}).
\ffigref{fig:linear}[D] shows the predicted and true shift vectors that generated the perturbation data.
The \oursanalysis shift predictions  $\widehat{\cb}_\pert$ follow directly from the matrix $\Ahat$ (Corollary \ref{corollary:clipr-vectors}).
The estimator precisely recovered the single-gene targets, the on-target perturbation effects, and the near-zero off-target effects.
The shifts $\widehat{\cb}_\pert$ were not constrained to be sparse or one-hot, but recovered the perturbation pattern solely from the inferred effects $\Ahat_\lambda$.

\begin{figure*}[t!]
    \centering
    \includegraphics[
        width=\textwidth,
        trim={1pt 67pt 1pt 1pt}, clip
    ]{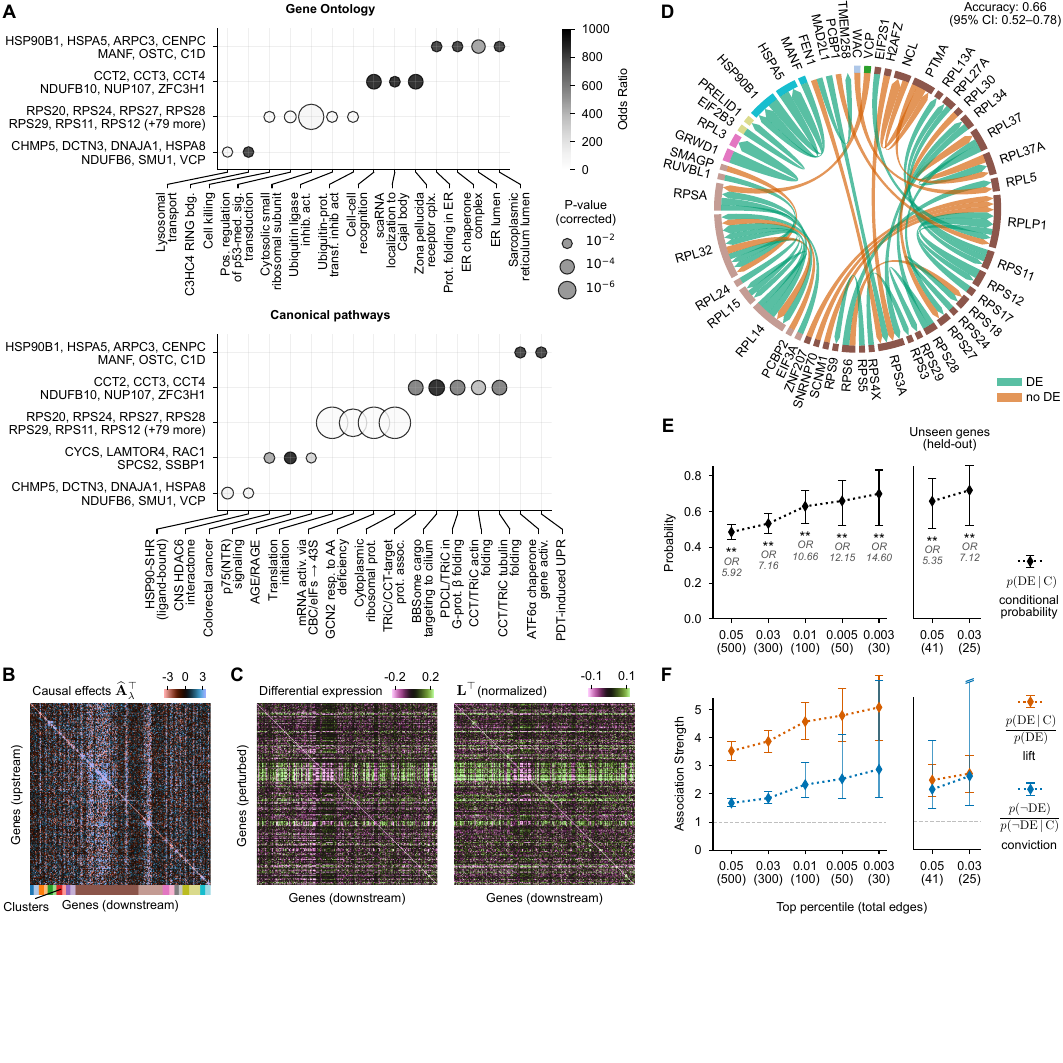}
    \caption{%
    \textbf{\ourscombo infers direct gene-gene causal effects from genome-wide \perturbseq data.}
	\subp{A}~Hierarchical clustering based on shortest-path distances in the \oursanalysis graph identifies gene modules enriched for specific Gene Ontology terms (top) and canonical pathways (bottom). 
	Clusters with at least one enriched term are shown.
	\subp{B} Predicted causal effects and associated clusters from the enrichment analysis. 
	\subp{C} Comparison between observed differential expression (\DE) (left) and normalized $\Finf^\top$ (right). 
	For visual clarity, only the 247 genes within identified clusters and their perturbations are shown; see \refappfig{fig:full-resolution-causal-replogle} and \ref{fig:full-resolution-causal-replogle-mats} for full matrices.
	\subp{D} Strongest predicted causal effects (50 total). 
	Edges are colored green where the predicted effect is validated by observed \DE, and red otherwise.
	(\subpp{E} and \subpp{F}) Validation of inferred causal links (\C) as a function of predicted effect strength.
	Stronger predicted causation significantly increases the probability of observing \DE (one-sided Fisher's exact test, $P < 0.01$, OR: odds ratio). 
	Results shown for all genes (left) and held-out genes not perturbed during training (right).
	Error bars indicate 95\% CIs (Wilson binomial for \subpp{E}, bootstrap for~\subpp{F}).
	}
    \label{fig:causality}
\end{figure*}

\subsection*{Disentangling causation from differential expression in Perturb-seq}

Our results suggest that \ourss accurately predict perturbation responses and recover the regulatory structure of testable linear systems.
We therefore reasoned that \ourscombo, now including the measurement model, can be used to infer {\em direct} causal effects among genes in \perturbseq screens.
This capability addresses a fundamental limitation of standard differential expression analyses,
which capture only {\em total} causal effects accumulated through the regulatory network.
This distinction matters: 
downstream expression changes may be mediated via intermediate transcription factors or signaling pathways rather than arising from direct regulation by the perturbed gene.
While perturbing gene $a$ may cause differential expression of gene $b$ when $a$ directly regulates $b$, the converse does not hold. 
Differential expression does not imply direct regulation, as the effect may be propagated through intermediate genes.
Understanding cellular regulatory programs therefore requires mapping direct causal mechanisms, not merely their downstream cascades.

We evaluated the capability of inferring the causal structure of direct gene-gene effects by training a \ours on a genome-wide \perturbseq assay of the human chronic myeloid leukemia (K562) cell line \citep{replogle2022mapping}.
This screen recorded single-gene perturbations of {\em all} genes modeled by the \ours, providing significantly more perturbation responses for \oursanalysis to estimate regulatory structure compared to the combinations datasets used for benchmarking (\ffigref{fig:benchmarking}; \refapptab{tab:datasets}).
While the perturbation dynamics learned by the \ours drift $f$ are likely highly nonlinear, our goal is to study the degree to which the linearized \oursanalysis causal effects predict plausible biological structure.

After estimating the \oursanalysis matrix $\Ahat_\lambda$ from the learned drift, we first performed hierarchical clustering of genes in the causal graph to study its network structure.
We grouped genes into clusters with strong average within-cluster causal effects (\methods).
To assess whether these clusters aligned with established biological annotations, we performed gene set enrichment analysis (GSEA) \citep{chen2013enrichr} using known ontology terms and canonical pathways \citep{liberzon2011molecular} \figref{fig:causality}[A].
Several clusters showed significant enrichment for both gene ontology terms and pathways (one-sided Fisher's exact test, $P < 0.01$, with Benjamini-Hochberg (BH) correction for multiple testing across gene clusters and annotations \citep{benjamini1995controlling}).
Moreover, clusters were enriched for distinct terms and pathways, suggesting that tight dependencies in the \ourscombo graph organize genes into functionally coherent modules. 
\ffigref{fig:causality}[B] shows the predicted causal effects $\Ahat_\lambda$ of the genes in the clusters analyzed for enrichment.
This matrix is qualitatively different than the genome-wide \DE matrix \figref{fig:causality}[C, left] when ordering the perturbed genes alongside the cause (upstream) genes.
Instead, the \ourscombo component $\Finf$ closely matches the differential expressions \figref{fig:causality}[C, right], since it captures the difference in the pre- and post-perturbation equilibria \figref{fig:method}[C]. 
Moreover, causal effects are, as expected, sparser than differential expressions when considering all $d=1000$ genes.
We provide plots of the full matrices, including $\Fzero$, $\Finf$, and $\Finf^+$, in \refappfig{fig:full-resolution-causal-replogle} and \ref{fig:full-resolution-causal-replogle-mats}.

Next, we validated the directed links predicted by \ourscombo.
In the absence of ground-truth annotations, our goal was to evaluate whether the implication discussed above---gene {\em a}  directly regulating gene {\em b} implying \DE of gene {\em b} when perturbing gene {\em a}---holds for the \oursanalysis estimates in the \perturbseq screen, as this implication can be assessed directly from the observed data.
Specifically, we studied how often this relationship holds as a function of the predicted effect strength, motivated by the observation that \ourscombo was most accurate for stronger predicted effects in simulated systems \figref{fig:linear}[B].
\ffigref{fig:causality}[D] shows the $50$ strongest predicted causal effects (top $0.005$ percentile of edges), many of which involved ribosomal protein (RP) genes,
and whether a perturbation of the upstream gene caused significant \DE downstream (Mann-Whitney $U$ Test, $P < 0.05$, BH corrected for multiple testing across genes).
To quantify significance of the implication, we computed the conditional probability of \DE given predicted causal dependence (\C) as well as
{\em lift} and {\em conviction}  \citep{brin1997dynamic}.
Lift measures the likelihood of \DE given \C compared to \DE unconditionally, and conviction quantifies how much less likely an implication violation (\C but not \DE) is given \C.
Similar to our results on simulated systems \figref{fig:linear}[B], stronger effects predicted by \ourscombo significantly increased the probability of observing \DE upon perturbation of the cause (Fisher's exact test, one-sided, $P < 0.01$) \figref{fig:causality}[E, left].
For the $50$ strongest predicted causal effects, observing significant \DE was approximately 4.78$\times$ more likely as measured by lift, and observing no \DE 2.53$\times$ less likely as measured by conviction, relative to the baseline \DE rates \figref{fig:causality}[F, left],
suggesting that strong \ourscombo links are consistent with the observed downstream effects.

\oursanalysis estimates a full causal structure $\Ahat_\lambda$ irrespective of whether all or only some genes were perturbed.
Hence, \ourscombo makes predictions about the causal effects of genes {\em not} originally perturbed  in the screen \figref{fig:linear}.
To test this generalization capability, we split the 41 upstream genes shown in \ffigref{fig:causality}[D] (all genes identified as causes among the top 50 strongest predictions) into five test folds and trained new \ours models, each leaving out perturbations for one fold.
We then applied \oursanalysis to each learned drift, re-aggregated the predicted causal effects for all held-out genes, and computed the previous implication metrics using the true held-out perturbation data.
Without observing any on-target perturbations, the strongest \ourscombo effects predicted for the held-out genes still significantly increased the probability of observing \DE when perturbing the cause gene, similar to when on-target perturbations were observed \figref{fig:causality}[E, right].
Significant \DE was approximately 2.71$\times$ more likely, and no \DE 2.62$\times$ less likely, relative to baseline \DE rates for top $0.03$ percentile of predicted edges (25 total) \figref{fig:causality}[F, right].

Overall, our findings suggest that \ourscombo can learn biologically plausible causal structure among genes, even when genes were not perturbed in the original screen.
Despite only approximating the perturbation behavior of the \ours drift, \oursanalysis organized genes into coherent functional clusters with tight causal effects and made meaningful predictions about which genes are affected by perturbations.
This highlights two main distinguishing factors between \oursanalysis and \DE analysis:
\oursanalysis infers direct causal effects, \DE only causal descendants;
and \oursanalysis can generalize to genes not perturbed in the original \perturbseq assay.

\section*{Discussion}

\Oursfulls were designed to address the need for perturbation models that are both predictive and interpretable \citep{rood2024toward,uhler2024building,tejada2025causal}.
As fundamentally {\em causal} models, they learn explicit, stochastic regulatory dynamics from \perturbseq count inputs, assuming a principled measurement model.
Perturbations modify the hidden states of the regulatory diffusion dynamics, and \oursanalysis enables interpreting these perturbed dynamics by constructing a linear approximation of the gene-gene causal structure of the entire system.
\ourss and \oursanalysis thus stand in contrast to state-of-the-art deep learning approaches, which are often black-box, deterministic, or based on prior gene relationship annotations \citep{lotfollahi2023predicting,bunne2023learning,roohani2024predicting,cui2024scgpt,he2025morph,he2025squidiff}.

Multiple temporal trajectories can generate the same aggregate population snapshots, so inference of causal dependencies from single-cell data has fundamental limits \citep{weinreb2018fundamental}.
Our approach alleviates this by learning from multiple snapshots, each modeled as a perturbation of the same causal regulatory dynamics.
Our results suggest that these learned dynamics often predict the outcomes of unseen perturbations more accurately than existing approaches \figref{fig:benchmarking}[B], while recovering causal dependencies consistent with ground-truth effects of simulated systems \figref{fig:linear} as well as known biology \figref{fig:causality}.
Theorem \ref{theorem:clipr} shows that snapshots of sufficiently diverse experimental conditions can provably identify causality, aligned with similar %
results showing that interventions improve identifiability 
\citep{peters2016causal,yang2018characterizing,zhang2023identifiability}.

The \ours model and \oursanalysis have potential limitations. %
In our experiments, \ourss performed best at predicting perturbation effects when the GI of a perturbation combination is unexpected, synergistic, or suppressive \figref{fig:benchmarking}[D].
When GI is additive or mean predictions suffice, simpler methods like SALT or PEPER \cite{gaudelet2024season}, which are less computationally intensive, perform competitively.
Similar to CPA, the \ours perturbation model relies on embeddings learned during training and thus only allows predicting the outcomes of combinations of observed perturbations \figref{fig:method}[B].
However, as \perturbseq now covers the full space of individual gene perturbations \citep{replogle2022mapping,carlson2023genome}, modeling their exponential number of combinations remains the key computational challenge, which \ourss enable us to do.
Beyond discovering GIs,  combinations are central to, for example, cell reprogramming \citep{wang2021direct} and drug combination therapies \citep{mokhtari2017combination}.
Nonetheless, \ourscombo predicted nontrivial causal structure even for unperturbed genes (\ffigref{fig:linear} and \ffigref{fig:causality}[E and F, right]), showing that \ourss can make inferences about unperturbed components of the system.
Ultimately, \oursanalysis remains a linear approximation of the learned  \ours drift and likely the true regulatory dynamics, which should be taken into account when interpreting the presence or absence of specific  links.
This may explain why larger \ourscombo effects tend to be more aligned with  differential expression (\ffigref{fig:linear}[B] and \ffigref{fig:causality}[E and F]).

Finally, \ourss and \oursanalysis have the prospect of integrating with different biological modalities and future modeling approaches.
We focused on gene perturbations to emphasize the difference between causality and differential expression,
but our framework directly extends to drug perturbation screens \citep{srivatsan2020massively}, since perturbation targets need not be known to learn \ourss or compute \oursanalysis.
Multiple batches or cell types may be integrated by conditioning the regulatory mechanisms on additional learned embeddings \figref{fig:method}[B], similar to foundation model approaches \citep{cui2024scgpt}.
Moreover, our method for inferring state densities from counts could extend to chromatin accessibility, surface protein readouts, or beyond transcriptomics.
Conversely, \ours drifts could be trained from gene states inferred by other denoising techniques \citep{van2018recovering}.
\oursanalysis could render the causal structure of any diffusion interpretable, provided it models perturbations, and therefore both \ourss and \oursanalysis may be of interest even outside of biology.
More generally, future research could explore approaches for merging the state density and drift inference steps without dropping established noise model assumptions or the separation of biological and measurement noise \figref{fig:method}[A].
Implementing these integrations would require calibrating assumptions made by various components and careful benchmarking with existing techniques.
Overall, stationary diffusions hold promise to complement \perturbseq screens with accurate learning-based prediction and to navigate the experimental design space by improving our causal understanding of the systems we assay.

\section*{\methodsname}

\phantomsection
\label{sec:methods}

\phantomsection
\subsection*{Inference}\label{sec:inference}

We first describe how to infer the generative model from data.
We model the discrete data distributions $p(\yb)$ as noisy observations of the stationary diffusion densities $p(\xb)$, sampled by the likelihood $p(\yb \given \xb; \pib)$.
Existing techniques for learning diffusions do not separate biological (diffusion) and measurement (technical) noise we distinguish here, so we infer the generative model in two steps \figref{fig:method}[A].
Suppose we observe data $\Yb_\pert = \{\yb^{(n)}\}$ for a control condition $\pert_0$ and $k$ perturbations $\pert_1, \dots, \pert_k$.
Inference then proceeds in two consecutive steps \figref{fig:method}[A]:
\begin{enumerate}
	\item\label{itm:inference-prior} 
	Learn likelihood parameters $\pib$ and state densities $p_\pert(\xb)$ 
	that maximize evidence of the observed~$\Yb_\pert$.
	\\
	\textbf{Input:}
	data $\Yb_\pert$ for each condition $\pert$
	\\
	\textbf{Output:}
	$\pib$ and samples $\Xb_\pert \sim p_\pert(\xb)$

	\item\label{itm:inference-diffusion}
	Learn drift $f$ and embeddings $\pertemb$ that jointly fit $p_\pert(\xb)$.
	\\
	\textbf{Input:}
	states $\Xb_\pert$ for each condition $\pert$
	\\
	\textbf{Output:}
	$f$ and embeddings $\pertemb$
\end{enumerate}
After both inference steps,
we have learned the stationary diffusion drift $f$, perturbation embeddings $\pertemb$, and measurement noise parameters $\pib$.
Together, these specify the full generative model and enable data generation under unseen embeddings $\pertembprime$.
The samples $\Xb_\pert$ are only used for fitting the drift (step \ref{itm:inference-diffusion}) and discarded afterwards. 
We now describe each step in more detail.

\paragraph*{Measurement model}
Inference of the state density $p(\xb)$ corresponds to learning the {\em prior} that generates the data distribution $p(\yb)$,
an approach called empirical Bayes \citep{robbins1956empirical}.
Our goal is not posterior inference of $p(\xb \given \yb)$, since we do not require the state that generated a specific datum $\yb$.
Our approach to this inference problem is to model the prior $p(\xb)$ as an implicit distribution that is easy to sample from and differentiate \citep{diggle1984monte}.
Specifically, we model prior samples $\xb \sim p(\xb; \phi)$ as $\xb = g(\epsilonb, \phi)$, where $\epsilonb$ is random noise and $g$ a differentiable function with respect to parameters $\phi$.
This choice enables a flexible distributional family for $p(\xb)$ to avoid biasing the drift trained on the inferred priors in the second inference step.
Implicitly modeling the prior is possible, because both of our inference steps only require samples $\xb \sim p(\xb; \phi)$ or differentiating them with respect to $\phi$.

We infer the prior $p(\xb; \phi_\pert)$ of a condition $\pert$ by maximizing the evidence of the observed readouts $\Yb_\pert$.
We use separate parameters $\phi_\pert$ for each condition $\pert$ but share the likelihood parameters $\pib$, as $\pib$ is used for test-time generation.
The evidence 
$\Zcal(\Yb_\pert, \phi_\pert, \pib)
= \log p(\Yb_\pert; \phi_\pert, \pib) $
is given by
\begin{align}
\begin{split}\label{eq:evidence}
	\Zcal
	(\Yb_\pert; \phi_\pert, \pib) 
	&= \sum_{n} \log \int  p(\xb;  \phi_\pert)p(\yb^{(n)} \given \xb; \pib) \mathrm{d}\xb 
    \\
	&\approx 
	\sum_{n}\logsumexp_{m=1}^M \left \{  \log p(\yb^{(n)} \given \xb^{(m)}; \pib)  \right \} - \log M \, ,
\end{split}
\end{align}
where we approximate the integral with $M$ samples
$\xb^{(m)} \sim p(\xb; \phi_\pert)$,
written with the log-sum-exp operator $\logsumexp_m \{z^{(m)}\} := \log\sum_m \exp(z^{(m)})$, which has a numerically stable implementation. 
The ZIP likelihood factors as 
$p(y_g \given x_g; \pi_g) 
= \pi_{g} \mathbb{I}(y_{g} = 0)
+ (1 - \pi_{g}) \mathrm{Pois}(y_g; \ratefunction_g(x_g))$,
where $\mathbb{I}(y_{g} = 0) = 1$ iff $y_{g} = 0$  \eqref{eq:noise-model}.
To account for scale differences across genes,
we used the rate function 
$\ratefunction_g(x_g) = \eta_g \log (1 + e^{x_g})$,
where $\eta_g$ is a fixed 
gene scaling
computed from control samples
(see {\secref{sec:training-details}{Training Details}}).

Optimization of $\Zcal$ is nontrivial, since the estimator \eqref{eq:evidence} is biased and high-variance.
Expanding 
$\smash{\log p(\yb \given \xb; \pib)} = \smash{\sum_{g=1}^d \log p(y_g \given x_g; \pi_g)}$
shows that the joint likelihood inside the log-sum-exp scales linearly with $d$.
Due to the exponential, maximizing the log-sum-exp term requires learning to generate samples with high likelihood across all $d$ dimensions jointly, which becomes exponentially less efficient with $d$.
We tackle this by using large $M$ and introducing a scaling $\tau^{-1}$ of $\log p(\yb \given \xb; \pib)$ in the loss:
\begin{align}
\begin{split}\label{eq:temperature-implicit-prior-loss}
	\widehat{\Zcal}_\tau(\Yb_\pert, \phi_\pert, \pib)
	&:= 
	\sum_{n} \tau \logsumexp_{m=1}^M \left \{ \tau^{-1}  \log p(\yb^{(n)} \given \xb^{(m)}; \pib)  \right \} \, .
\end{split}
\end{align}
The temperature $\tau$ interpolates between
fitting the full joint distribution including higher-order dependencies ($\tau = 1$, higher variance)
and fitting the marginal distributions only (high $\tau$, lower variance).
As $\tau \rightarrow \infty$, the log-sum-exp operator converges to the arithmetic mean,
effectively swapping logarithm and summation like a Jensen's lower bound,
and treating the $d$ dimensions in the joint likelihood 
as (unconditionally) independent, thus alleviating the variance caused by the exponential.
Our final training objective $\Lcal_{\mathrm{IP}}$ for learning the implicit prior is to minimize the combined loss
\begin{align}\label{eq:implicit-prior-loss}
\Lcal_{\mathrm{IP}} 
= - \widehat{\Zcal}_1 
- \widehat{\Zcal}_\tau
\end{align}
tuned over $\tau \in \{5, 20, 100\}$.
$\Lcal_{\mathrm{IP}}$ is fully differentiable with respect to $\phi$ and $\pib$ by propagating gradients through the implicit samples $\xb = g(\epsilonb, \phi)$.

\paragraph*{Stationary diffusion}
Inferring the drift $f$ and perturbation embeddings $\pertemb$ requires learning a stationary diffusion model that jointly fits the densities $p_\pert(\xb)$ obtained in step \ref{itm:inference-prior}.
Any learning approach may in principle be used for this.
In our experiments on SDEs with constant $\sigma$ \eqref{eq:causal-model}, score matching was computationally efficient and achieved good fit and generalization.
In this setting, a drift can be inferred by denoising smoothed data $\xbc \sim \Ncal(\xbc; \xb, \dsmscale^2\Ib)$, 
where $\xb \sim p(\xb)$.
The drift $f$ that minimizes the squared loss $\smash{\lVert f(\xbc) + (\xbc - \xb)/\dsmscale^2 \rVert^2_2}$ in expectation
satisfies $f(\xbc) = \nabla \log p(\xbc; \dsmscale)$ under weak regularity conditions, which implies the SDE has stationary density $\smash{p(\xbc; \dsmscale)}$ for $\smash{\sigma = \sqrt{2}}$  \citep{vincent2011connection}.
The difference between $\nabla\log p(\xbc; \dsmscale)$ and $\nabla \log p(\xb)$ becomes negligible as $\dsmscale \rightarrow 0$.
Following Song and Ermon \citep{song2019generative},
we thus infer $f$ by learning to denoise different smoothing levels $\dsmscale$, conditioning the drift on $\dsmscale$.
The loss for drift $f(\xb; \theta, \eb_\pert, \dsmscale)$ is then
\begin{align}
\begin{split}\label{eq:drift-loss}
    \hspace{-3pt}
	\Lcal_{\mathrm{D}}(\Xb_\pert, \theta, \eb_\pert, \dsmscale^{(n)})
	\hspace{-1pt}
	&=
	\hspace{-1pt}
	\sum_{n}
	\hspace{-1pt}
	\big \lVert
	\dsmscale^{(n)}
	f\big(
	\widetilde{\xb}^{(n)};
\theta, \eb_\pert, \dsmscale^{(n)}
	\big)
	\hspace{-1pt}
	+ \epsilonb^{(n)}
	\big\rVert_2^2
\end{split}
\end{align}
where 
$\widetilde{\xb}^{(n)} =
\xb^{(n)} \hspace{-1pt}
	+ \dsmscale^{(n)}\epsilonb^{(n)}$,
$\epsilonb^{(n)} \sim \Ncal(\bzero,\Ib_d)$,
and the scales
$\dsmscale^{(n)}$ are drawn from the geometric sequence $\Dsmscale = [\num{0.01}, \dots, \num{10}] \in \RR^{30}$.
In Eq.~\ref{eq:drift-loss}, we scaled the norm by $\dsmscale^2$ to ensure that loss terms have comparable magnitudes.
Training on multiple and large smoothing levels $\dsmscale$ avoided a behavior we called `multi-fitting':
without higher smoothing scales $\dsmscale$, diffusions tended to fit all perturbed densities $p_\pert(\xb)$ unconditionally like a mixture distribution of all conditions, ignoring the embeddings $\pertemb$.
Avoiding this also required careful sampling based on $\Dsmscale$
(see {\secref{sec:sampling}{Sampling}}).
For \oursanalysis, we condition $f$ on the smallest $\dsmscale_1 = 0.01$ and thus least smoothing error
(see {\secref{sec:extended-clipr}{\oursanalysis}}).
We fixed the diffusion scale $\sigma$ \eqref{eq:causal-model} to $\smash{\sigma = \sqrt{2}}$  throughout all experiments
and shared $\theta$ across conditions $\pert$, modeling the gene regulatory mechanisms and enabling test-time generation for unseen $\eb_{\pert}$.

Score matching fits the drift $f(\xb; \theta, \eb_\pert)$ to a gradient field, %
abstracting away the fact that $\theta$ is fit to multiple densities and coupled through $\eb_\pert$.
While this would imply that steady-state dynamics have zero curl ($\partial [f_\pert]_a/\partial x_b = \partial [f_\pert]_b/\partial x_a$), the perturbation behavior modeled by \ourss is not symmetric. 
Because perturbations modify the hidden state of a nonlinear drift \figref{fig:method}[B],
perturbing gene $a$ may not have the same effect on gene $b$ as perturbing gene $b$ has on gene $a$.
\oursanalysis reveals that \ours perturbations indeed exhibit asymmetry (\ffigref{fig:linear}[C] and \ffigref{fig:causality}[B]).
To learn stationary densities with non-zero curl and enable modeling, for example, negative feedback loops of three repressors 
\citep{elowitz2000synthetic,weinreb2018fundamental},
future work could leverage more general learning objectives not based on the gradient field assumption \citep{lorch2024causal}.

\phantomsection
\subsection*{Sampling}\label{sec:sampling}

To generate data $\yb \sim p_{\pert}(\yb)$ for a test perturbation $\eb_{\pert}$,
we conditioned $f$ on the test embedding $\eb_{\pert}$ and simulated the diffusion to obtain samples $\xb \sim  p_{\pert}(\xb)$ from the stationary density.
For a perturbation combination of $\pert_a$ and $\pert_b$, the embedding is $\eb_{\pert_a} + \eb_{\pert_b}$ (see {\secref{sec:neural-network-architectures}{Neural network architectures}}). 
We generated diffusion samples by annealing the Euler-Maruyama method according to the trained smoothing levels $\dsmscale$ \citep{song2019generative}.
The sample path approximation at step $l$ is \citep{sarkka2019applied}
\begin{align}
\begin{split}
\label{eq:euler-approximation}
    \xb^{(l+1)}
    =
    \xb^{(l)}
    + f\big(\xb^{(l)}; \theta, \eb_{\pert}, \dsmscale^{(l)} \big) 
    (\dsmscale^{(l)})^2
    \Delta t 
    + \xib^{(l)}
    \dsmscale^{(l)}
    \sqrt{2 \Delta t}
\end{split}
\end{align}
for noise $\xib^{(l)} \sim \Ncal(\bzero,\Ib_d)$,
step size $\Delta t$,
and smoothing level $\dsmscale^{(l)}$.
In our experiments, 
we used 
$\Delta t = 10^{-4}$,
initialized at $\xb(0) \sim \Ncal(\bzero, \Ib_d)$,
and computed \num{500} 
Euler-Maruyama steps %
per annealing scale in $\Dsmscale$, 
starting with $\alpha_{30} = 10.0$ and ending at $\alpha_{1} = 0.01$.
The last state $\smash{\xb^{(l)}}$ corresponds to the sample from $p_{\pert}(\xb)$
after inverting the shift and scale standardization performed
on the training data (see {\secref{sec:training-details}{Training details}}).
For each $\xb \sim  p_{\pert}(\xb)$,
we generated one observation $\yb \sim p(\yb \given \xb; \pib)$ \eqref{eq:noise-model} as the final sample from the \ours.

\phantomsection
\subsection*{Neural network architectures}\label{sec:neural-network-architectures}
For our experiments, we parameterized the drift and the implicit priors using multi-layer perceptrons (MLPs).
The implicit prior network $g(\epsilonb, \phi)$
transformed noise $\epsilonb \sim \Ncal(\epsilonb; \bzero, \Ib_{64}) \in \RR^{64} $ into  samples $\xb = g(\epsilonb, \phi) \in \RR^{d}$.
After an initial linear layer mapping $\epsilonb$ to \num{64} dimensions,
the network consisted of a sequence of 
residual MLP blocks with \num{64} hidden dimensions (\num{4} blocks),
followed by a linear layer mapping to $d$ dimensions,
and additional residual MLP layers with \num{256} hidden dimensions (\num{2} blocks).
Residual MLP blocks computed output $\zb'$ from input $\zb$ as
$\zb' := \zb + \mathrm{Linear}(\mathrm{SiLU}(\mathrm{Linear}(\zb)))$,
where $\mathrm{SiLU}(\zb) := \zb * \mathrm{sigmoid}(\zb)$ elementwise
and $\mathrm{Linear}(\zb) := \Wb \zb + \bb$
with independent parameters $\Wb,\bb$ and corresponding dimensions per layer.
Residual layers stabilized training and the variance of $p(\xb; \phi)$,
and the low dimensionality reduced memory, enabling larger Monte Carlo sample sizes $M$ during training.

The drift $f(\xb; \theta, \eb_\pert, \dsmscale)$ was parameterized by a fixed $\dsmscale$-dependent scaling to adjust for changing state and gradient magnitudes across the different smoothing scales $\dsmscale$ \citep{song2021scorebased}.
Specifically, we modeled $f$ as
\begin{align}
	f(\xb; \theta, \eb_\pert, \dsmscale)
	&=
	\frac{1}{\dsmscale}
	h
	\bigg(
	\frac{\xb}{\sqrt{1 + \dsmscale^2}}; 
	\theta, \eb_\pert, \dsmscale
	\bigg)
\end{align}
where $h$ is the main neural network parameterized by $\theta$, taking state $\xb$, perturbation embedding $\pertemb$, and smoothing scale $\dsmscale$ as input.
$h$ is an MLP with two hidden layers, $\tanh$ nonlinearities, and $\dimdrift \in \{\num{1024}, \num{2048}\}$ dimensions  \figref{fig:method}[B].
Representations $\pertemb$ have \num{1024} dimensions and are jointly learned with $\theta$.
They are added to both hidden states of the MLP before the nonlinearity following separate learned linear transformations,
similar to conditioning techniques in vision models 
\citep{dumoulin2017learned}.
We model combinations of perturbations $\pert_a$ and $\pert_b$ by adding embeddings as
$\eb_{\pert_a, \pert_b} = \eb_{\pert_a} + \eb_{\pert_b}$ and represent the control condition $\pert_0$ by the zero vector $\eb_{\pert_0} = \bzero$.
Lastly, $h$ models the effect of $\dsmscale$ 
by encoding the index of $\dsmscale$ in the sequence $\Dsmscale$
as a sinusoidal position embedding \citep{vaswani2017attention},
then mapping to $\dimdrift$ dimensions with a linear layer, $\tanh$ nonlinearity, and a second linear layer,
and finally adding the $\dsmscale$ representation to the first hidden state of~$h$.

\phantomsection
\subsection*{Training details}\label{sec:training-details}

As described in {\secref{sec:inference}{Inference}},
we observe data $\Yb_\pert$ for a control condition $\pert_0$ and $k$ perturbations $\pert_1, \dots, \pert_k$. 
Training then proceeds in two consecutive steps.
We used the same training hyperparameters devised during development of the method across all experiments unless we note they are tuned otherwise  (see {\secref{sec:baselines-hyperparameters}{Baselines and hyperparameters}}).

We learned the priors by minimizing 
$\Lcal_{\mathrm{IP}}$ \eqref{eq:implicit-prior-loss} with gradient descent 
for \num{250000} steps.
We used the Adam optimizer \citep{kingma2015adam} 
($\beta_1 = 0$, $\beta_2 = 0.99$)
with learning rate $0.0003$,
warmed up for \num{5000} steps and cosine-decayed to zero.
Each step randomly sampled a perturbation $\pert$ and
then updated parameters $\phi_\pert$ and $\pib$ based on $\Yb_\pert$. 
The loss $\Lcal_{\mathrm{IP}}$ was computed using a batch size of $64$ and $M=\num{20000}$ Monte Carlo samples.
Gradients were clipped at $\ell_2$-norm $1.0$.
We fixed the gene-specific scalings in the Poisson rate $\ratefunction_g(x_g)$ \eqref{eq:noise-model} to
$\smash{\eta_g = \overline{y}_g + \overline{s}^2_g/\overline{y}_g - 1}$, 
lower bounded by $1$,
where $\smash{\overline{y}_g}$ and $\smash{\overline{s}^2_g}$ are sample means and variances of $y_g$ in the control data $\Yb_{\pert_0}$.
This is the method-of-moments estimator for the Poisson mean under a ZIP model of $y_g$ \citep{beckett2014zero} and thus calibrated $\ratefunction_g(x_g)$ to the expected scale of $y_g$.
Training the implicit priors did not overfit, since the Poisson has a minimum level of dispersion, and therefore training did not include explicit regularization.
We validated model fit by tracking the evidence $\Zcal$ and comparing generated samples to $\Yb_\pert$ after passing them through the learned likelihood.
After optimization, we generated $N=\num{10000}$ state samples 
$\Xb_\pert = \{\xb^{(n)}\}_{n=1}^{N} \sim p(\xb; \phi_\pert)$ for training the diffusion.

We trained the drift $f$ and perturbation embeddings $\pertemb$ 
by minimizing $\Lcal_\mathrm{D}$ \eqref{eq:drift-loss}
with gradient descent
for \num{50000} steps.
We used the Adam optimizer ($\beta_1 = 0.9$, $\beta_2 = 0.999$)
with learning rate \num{0.001},
cosine-decayed to zero,
and weight decay tuned over
$\{0.03, 0.1, 0.3, 1.0, 3.0\}$.
Weight decay was most effective at controlling over- and underfitting and tuning predictive performance of \ourss.
As during prior inference (step \ref{itm:inference-prior}),
each update step randomly selected one condition $\pert$, one smoothing level $\alpha$, and then updated parameters $\theta$ and $\pertemb$,
using a batch size of \num{8192} to compute $\Lcal_{\mathrm{D}}$.
We standardized the training data  
to facilitate gradient-based optimization
by subtracting the control means of $\Xb_0$
and dividing by the standard deviations across all training samples $\Xb_0, \dots, \Xb_k$.
\refapp summarizes the computational requirements of training.

\phantomsection
\subsection*{\oursanalysis}\label{sec:extended-clipr}

We numerically approximated the fixed point defining the perturbation response $\finfpert$ by integrating the flow of $f$ conditioned on $\pertemb$ as
\begin{align}\label{eq:finf-approx}
	\finfpert 
	&\approx
	\lim_{t\rightarrow \infty} \overline{\xb}(t)\, , 
\end{align}
where
$\dd\overline{\xb}(t)/\dd t = f(\overline{\xb}(t); \theta, \eb_\pert, \alpha_1)$ and $\overline{\xb}(0) = \bzero$
\figref{fig:method}[C].
For integration, we used the same annealing scheme as used for generation but dropped the noise term
(see {\secref{sec:sampling}{Sampling}).
Integration starting from the control vector $\bzero$ resolves the ambiguity that $f_\pert$ may admit multiple fixed points.
We assume throughout that this limit exists, which held true in all experiments.
To compute \oursanalysis, we assembled the 
$\Fzero$ and $\Finf$ matrices (Theorem \ref{theorem:clipr})
by conditioning the drift $f$ separately on each perturbation embedding $\pertemb$ learned during training
(i.e., ignoring combinations, which are modeled as sums of embeddings).
We used a regularization strength of $\lambda = 1/10d$ for systems of $d$ variables,
which worked well across all settings (Figs.~\ref{fig:linear} and \ref{fig:causality}),
specifically
$\lambda = 10^{-3}$ for $d=100$,
$\lambda = 3 \cdot 10^{-3}$ for $d=300$,
and $\lambda = 10^{-4}$ for $d=1000$.

We standardize the states $p(\xb)$ before learning $f$,
so the learned $f$ operates in different unit scales than the original $p(\xb)$ 
(see {\secref{sec:training-details}{Training Details}}).
However, we still want to interpret the causal effects of the drift corresponding to the original units when applying \oursanalysis to $f$ learned in standardized units.
Put differently,
we seek the \oursanalysis linearization of the drift we {\em would have learned} had the data not be standardized.
To accomplish this, additional results in \refapp  %
provide the original drift $f^x$ as an explicit function of the data transformation and learned drift $f^s$ assuming the diffusion scale $\sigma$ is fixed.
We show that,
if densities $p(\xb)$ are transformed such that $\xb = \tbias + \tmat\sbb$, where $\tbias \in \RR^d$ and $\tmat \in \RR^{d \times d}$ is positive definite,
the perturbation responses of $f^x(\xb)$ 
(Eqs.~\ref{eq:fzero-vector} and \ref{eq:finf-vector})
are given
in terms of 
$f^s(\sbb)$ 
as
\begin{align}
\begin{split}\label{eq:perturbation-responses-transformed-brief}
	\fzero_{f^x}
    &= \tmat^{-1} f^s (-\tmat^{-1} \tbias) \, , \\
	\finf_{f^x}
	&=  \tbias + \tmat\finf_{f^s} \, .
\end{split}
\end{align}
We always applied Eq.~\ref{eq:perturbation-responses-transformed-brief} before computing \oursanalysis to account for data standardization before training.
\refapp provides formal statements of these results and applies them to linear SDEs as a practical illustration.

\phantomsection
\subsection*{Datasets and preprocessing}\label{sec:datasets-preprocessing}

We performed experiments on \perturbseq data by Norman et al.\ \citep{norman2019exploring} and Wessels et al.\ \citep{wessels2023efficient} (\ffigref{fig:benchmarking})
and by Replogle et al.\ \citep{replogle2022mapping} (\ffigref{fig:causality}).
All datasets were prepared with standard preprocessing steps.
Initial quality control filtered 
genes expressed in less than \num{20} cells,
cells measuring less than \num{200} genes,
and perturbations with less than \num{100} cells,
as well as outliers based on
dataset-specific cutoffs for high mitochondrial RNA counts
and total cell and gene counts.
For benchmarking, we split the double-gene perturbations into a validation set of \num{20} perturbations for hyperparameter tuning and a test set used for benchmarking, which was split further into \num{10} folds.
For a given test fold, methods were trained on control, single-gene, and double-gene perturbations of the remaining test folds.
Combining all experiments, we obtained test-time predictions for all double-gene perturbations (\refappfig{fig:illustration-benchmark-splits}).
\refapptab{tab:datasets} lists summary statistics for the perturbations available in each dataset.

For all datasets, we modeled \num{1000} highly variable genes 
based on Seurat-dispersions \citep{satija2015spatial} of the control and single-gene perturbation data (always in training split),
normalized to median control library sizes and $\logp$ transformed.
Gene selection prioritized perturbed genes and top \num{20} marker genes of the perturbations.
Norman et al.\ recorded the single-gene perturbations underlying most double-gene perturbations, 
which enabled us to classify gene pairs into GI categories by comparing the single and double post-perturbation means (\refapp).
GI metrics computed on the \num{1000} selected genes correlated with those computed on the full gene set
(\refappfig{fig:comparison-gi-metrics}).
We therefore determined thresholds for GI types based on the classifications of the original study
(\refappfig{fig:comparison-gi-thresholds}).

We also performed experiments on perturbation data generated by linear systems
(Fig.~\ref{fig:linear}).
For each evaluation setting, we randomly generated ten linear systems \eqref{eq:linear-sde}
with different parameters $\Ab, \bb$ 
and then sampled single-gene perturbation data using one-sparse shift vectors $\cb_\pert$.
Specifically, we sampled $\Ab\sim \unifpm(1, 3)$ and $\bb \sim \unif(-3, 3)$,
where $\unifpm$ denotes a uniform distribution with random signs,
masked $\Ab$ by random \erdosrenyi and power-law sparsity structures,
and then shifted the masked $\Ab$ by a constant diagonal to ensure stability with a maximum eigenvalue of \num{-0.5} \citep{lorch2024causal}.
\erdosrenyi structures contained independently-sampled non-zero entries (\num{10} per gene in expectation) \citep{erdos1959random},
and power-law structures were randomly generated with few regulator genes and many downstream genes
(Barab{\'a}si-Albert preferential attachment with strength \num{2} and \num{10} links per gene, orientation flipped with probability \num{0.1} to create feedback loops) \citep{barabasi1999emergence}.
Each perturbation $\cb_\pert$ targeted a unique gene $g$ with on-target shift $[\cb_\pert]_g \sim \unifpm(5, 15)$ and zero entries otherwise.
The simulated datasets contained \num{200} samples per perturbation.

\phantomsection
\subsection*{Enrichment Analysis}\label{sec:perturb-seq-clipr}
We analyzed the causal structure inferred from the \perturbseq screen  \figref{fig:causality}[A] by hierarchically clustering genes based on their causal effects. 
Gene distances were defined as shortest path lengths in the graph 
$\smash{\Wb_\lambda + \Wb_\lambda^\top}$, 
where $\Wb_\lambda = 1/ \lvert \Ahat_\lambda + 10^{-10} \rvert $
to ensure symmetry and capture indirect dependency.
We then clustered genes by iteratively linking sets with minimal average distances.
We obtained the final set of tight gene clusters by cutting the dendrogram at the lowest height that still grouped ribosomal protein genes into a single cluster,
which constituted a large, tight cluster upon visual inspection (\refappfig{fig:full-resolution-causal-replogle}).
Clusters containing fewer than five genes as well as genes not assigned to any cluster after cutting the dendrogram were excluded from further analysis.
We performed GSEA on the predicted gene clusters 
by testing for overrepresentation in
gene ontology terms (C5.GO)
and canonical pathways (C2.CP)
of the MSigDB database \citep{liberzon2011molecular},
using one-sided Fisher's exact tests 
and correcting for multiple testing across both gene clusters and annotations.
\ffigref{fig:causality}[A] shows top five enriched terms per predicted cluster, ranked by odds ratio.
We increased the \oursanalysis regularization to $\lambda = 10^{-2}$ compared to the benchmarking datasets, which included either substantially fewer perturbations or no model mismatch, as the matrix $\Ahat_\lambda$ had not yet stabilized at the default $\lambda = 10^{-4}$ 
(see {\secref{sec:extended-clipr}{\oursanalysis}}).

\phantomsection
\subsection*{Metrics}\label{sec:metrics}

For benchmarking, each method generated $\num{1000}$ samples per test perturbation \figref{fig:benchmarking}.
We computed MMD using a squared exponential kernel
$k(\yb, \yb') = \exp(- \lVert \yb - \yb' \rVert_2^2/2d\gamma^2)$,
averaged over \num{10} length scales $\gamma \in [\num{0.01}, \num{10}]$ in geometric progression \citep{bunne2023learning}.
We also calculated RMSE and Pearson correlation between the predicted and observed means, respectively, aligned with previous works \citep{lotfollahi2023predicting,roohani2024predicting}.
We identified the top \num{20} differentially expressed (marker) genes relative to control using Mann-Whitney $U$ tests with BH correction.
Metrics were computed on normalized counts (library size $10^4$, $\logp$ transformed).
An explicit definition of the empirical MMD is given in \refapp.

For evaluating \oursanalysis, we computed the accuracy of $\Ahat_\lambda$ relative to the ground-truth $\Ab$ by classifying the predicted effects into positive, negative, and absent using the thresholds $\pm\num{0.5}$ \figref{fig:linear}.
We reported one-versus-one AUROC scores averaged over the three class pairs, using predicted effect strength as confidences, which is insensitive to class imbalance \citep{hand2001simple}.
We calculated the F1 score for the three-class setting by averaging the scores predicted for each label.
All metrics excluded the diagonals of the causal effect matrix (self-links).

\phantomsection
\subsection*{Baselines and hyperparameters}\label{sec:baselines-hyperparameters}

We compared \ourss to various methods  \figref{fig:benchmarking}.
All figures show the baseline performance achieved by random control and perturbed cells of all training perturbations.
SALT predicted post-perturbation effects of double perturbations as the sum of their constituent single perturbation effects \citep{gaudelet2024season}.
Specifically, let $\Delta_\pert \in \RR^d$ denote the mean differential expression of the normalized and log-transformed counts after perturbation $\pert$.
Then, SALT generated samples for a double perturbation $\pert_a, \pert_b$ by shifting randomly sampled control data by $\Delta_{\pert_a} + \Delta_{\pert_b}$.
If $\pert$ was not observed as a single perturbation, SALT imputed the mean across all training perturbations.
\refapp provides details on the other baselines.

We tuned loss temperature $\tau$, hidden size $d_f$, and weight decay for \ourss and key hyperparameters of each baseline.
For this, we selected configurations that minimized the median MMD (top \num{20} DE genes) evaluated on a validation split of the double-gene perturbations (\refappfig{fig:illustration-benchmark-splits}).
\secref{sec:training-details}{Training Details} provides details for \ours and \refapp for the baselines.
For the linear systems and genome-wide \perturbseq and experiments (Figs.~\ref{fig:linear} and \ref{fig:causality}), 
we trained on significantly more perturbations and thus doubled the drift training steps. %
Because these datasets lack double perturbations for validation, we fixed hyperparameters at intermediate values of $\tau = \num{20}$, weight decay $=$ \num{1}, and $d_f = \num{2048}$.

\section*{Acknowledgments and Disclosure of Funding}
We thank Elvira Forte, Frederike L{\"u}beck, and Scott Sussex for helpful feedback on the manuscript.
This work has also greatly benefited from discussions with Marco Bagatella, Frederike L{\"u}beck, Matteo Pariset, Scott Sussex, Lenart Treven, and fellows of the Eric and Wendy Schmidt Center at the Broad Institute.
L.L.\ and A.K.\ were partially supported by the Swiss National Science Foundation under NCCR Automation, grant agreement 51NF40 180545 and L.L.\ and B.S.\ were partially supported by the Deutsche Forschungsgemeinschaft (DFG, German Research Foundation) under Germany’s Excellence Strategy -- EXC number 2064/1.
J.Z.\ was supported by the Eric and Wendy Schmidt Center at the Broad Institute.
C.U.\ was partially supported by NCCIH/NIH (1DP2AT012345), NIDDK/NIH (5RC2DK135492-02), ONR (N00014-24-1-2687), and the United States Department of Energy (DE-SC0023187).

\section*{Author Contributions}
L.L., J.Z., C.B., A.K., B.S., and C.U. designed research;
L.L., J.Z., and C.B. conceived the LCD model;
L.L. developed inference steps, CLIPR, software, and performed experiments;
L.L., J.Z., C.B., A.K., B.S., and C.U. analyzed data and wrote the paper;
A.K., B.S., and C.U. supervised the research.

\section*{Declaration of Interests}
The authors declare no competing financial interests.

\section*{Data and Code Availability}
All code was implemented in Python. 
The source code will be released upon publication of this work.

\makeatletter
\renewcommand\@biblabel[1]{#1.} %
\makeatother

{\footnotesize
\bibliographystyle{unsrtnat}
\bibliography{ref.bib}
}

\newpage
\clearpage

\renewcommand{\thefigure}{S\arabic{figure}}
\setcounter{figure}{0}
\renewcommand{\thetable}{S\arabic{table}}
\setcounter{table}{0}

\section*{\Large Supplementary Information}
\setcounter{theorem}{2}
\setcounter{equation}{16}
\setlength{\skip\footins}{15pt}

\phantomsection\label{sec:appendix}

\medskip

\section*{CLIPR Under Change of Variables}

When performing a variable transformation of the states $\xb$ before learning a drift $f$, learned drifts will operate in different unit scales than the original state distribution $p(\xb)$.
Our goal is to obtain \oursanalysis estimates for $f$ in original units given only access to the transformation parameters and the learned $f$.
To accomplish this, the following theorem provides the original drift  as an explicit function of the transformation and the learned drift, assuming the diffusion scale is fixed:
 \begin{theorem}\label{theorem:drift-transformed}
Let $f^s(\sbb)= -\nabla u(\sbb)$ be the drift of an SDE 
$\mathrm{d}\sbb(t) = f^s(\sbb(t))\mathrm{d}t + \sigma \mathrm{d}\mathbb{W}(t)$
with $u: \RR^d \rightarrow \RR$ 
that induces stationary distribution $p(\sbb)$.
Suppose the variables are transformed as
$\xb = \tbias + \tmat\sbb$, where $\tbias \in \RR^d$ and $\tmat \in \RR^{d \times d}$ is positive definite.
Then, the drift $f^x(\xb)$ of an SDE that induces the stationary distribution $p(\xb)$ with the same diffusion scale $\sigma$ is 
\begin{align*}
    f^x(\xb) 
    = \tmat^{-1} f^s \big(\tmat^{-1}(\xb - \tbias)\big) \, .
\end{align*}
 \smallskip
 \end{theorem}

Theorem \ref{theorem:drift-transformed} derives the drift of an SDE with the same diffusion scale $\sigma$ as the original SDE, because $\sigma$ is typically not changed based on a priori transformations such as data standardization.
Without fixing $\sigma$, 
the transformed drift follows directly from It{\^o}'s Lemma \citep{oksendal2003stochastic} 
and restrictions on $f$ and the transform can be relaxed.
The gradient field and integrability condition ensure that $f$ induces a stationary distribution and has zero curl.
Since we train the drift $f$ with score matching,
which approximately yields a gradient field, 
Theorem \ref{theorem:drift-transformed} can be used for analyzing \ours drifts despite the standardization used in training.
The following corollary gives the perturbation responses of the drift $f^x$ in Theorem \ref{theorem:drift-transformed} as an explicit function of $f^s$ and the parameters of the transform:

 \begin{corollary}\label{corollary:clipr-vectors-transformed}
The perturbation responses 
of $f^x$ in Theorem \ref{theorem:drift-transformed} can be expressed in terms of $f^s$ as
\begin{align*}
	\fzero_{f^x}
    &= \tmat^{-1} f^s (-\tmat^{-1} \tbias) \, , \\
	\finf_{f^x}
	&=  \tbias + \tmat\finf_{f^s} \, .
\end{align*}
 \end{corollary}

\section*{Proof of Theorem \ref{theorem:clipr}}

We seek the parameter matrix $\Ab$ that induces the perturbation response vectors
$\smash{\{\fzeropertstart, \dots, \fzeropertend\}}$ and 
$\smash{\{\finfpertstart, \dots, \finfpertend\}}$
under the linear model 
\eqref{eq:linear-sde}
with the least squared approximation error and minimum Frobenius norm.

By definition of $\flin_\pert$,
it must hold that 
$\fzeropert = \bb + \cb_\pert$ for the linear model to induce the initial perturbation response $\fzeropert$.
Moreover, 
since the post-perturbation mean $\mub_\pert$ is the unique fixed point of the linear drift,
the linear model must satisfy $\finfpert = \mub_\pert$.
Substituting both conditions into the analytical form of the mean 
\eqref{eq:linear-sde-mean},
we obtain the equality $\Ab \finfpert = - \fzeropert$.
Combining the conditions for all perturbations $\{\pert_1, \dots, \pert_k\}$, we obtain the system of linear equations
\begin{align}\label{eq:clipr-proof-system}
	\Ab \Finf = - \Fzero	\, ,
\end{align}
where $\Fzero$ and $\Finf$ are defined as in the theorem statement.
If $\Ab$ satisfies the equality \eqref{eq:clipr-proof-system}, 
the linear SDE model with causal effects $\Ab$ induces exactly the perturbation responses specified by $\Fzero$ and $\Finf$.

In general, this system may, however, be underdetermined (infinitely many solutions $\Ab$) or inconsistent (no solution $\Ab$).
We therefore seek the least squares solution $\Ahat = \arg \min_\Ab \lVert \Ab \Finf + \Fzero\rVert_\mathrm{F}^2$
with minimum norm $\lVert \Ab \rVert_\mathrm{F}^2$,
which always exists, is unique, and is given by the Moore-Penrose pseudoinverse \citep{penrose1956best}
\begin{align*}
	\Ahat = - \Fzero  \Finf^+	\, .
\end{align*}
Moreover, if $\Finf \in \RR^{d \times k}$ has full row rank, 
$\Ahat$ is the only least squared solution \citep{golub2013matrix}.
Thus, if additionally an exact solution $\Ab$ of Eq.~\ref{eq:clipr-proof-system} exists,
where the least squared error is minimized at zero,
the matrix $\Ahat$ uniquely identifies this solution.
This implies that $\Ahat = \Ab$ if the perturbation response matrices $\Fzero$ and $\Finf$ are generated by a linear drift $\smash{f = \flin}$, where an exact solution exists (the true $\Ab$),
and $\Finf$ has full row rank.
A sufficient condition for $\Finf$ being full row rank is that $k \geq d$ perturbation shift vectors $\cb_{\pert_i}$ are linearly independent.
This fails only in the degenerate case where $\bb$ is an an affine combination of the perturbation shifts, i.e.,
$\smash{\bb = \sum_{i=1}^k \beta_i \cb_{\pert_i}}$ with $\smash{\sum_{i=1}^k \beta_i = 1}$.
Otherwise,
the vectors $\bb + \cb_{\pert_i} = \finfperti$ are linearly independent, so $\Finf$ has full  rank.

When Eq.~\ref{eq:clipr-proof-system} is inconsistent, 
the least squares solution $\Ahat$ may not be optimal in all applications.
The estimator can be made more well-behaved by regularizing the norm of the solution itself via
$\Ahat_\lambda = \arg \min_\Ab \lVert \Ab \Finf + \Fzero\rVert_2^2 + \lambda \lVert \Ab \rVert_2^2$, known as Tikhonov regularization.
The analytical solution to this regularized problem is given by \citep{tikhonov1977solutions}
\begin{align*}
	\Ahat_\lambda =
    - \Fzero 
    \Finf^\top
    \big (
    	\Finf\Finf^\top 
    	+ \lambda \Ib
    \big )^{-1}
    \, . 
\end{align*}

\hfill $\blacksquare$

\section*{Proof of Corollary \ref{corollary:clipr-vectors}}

The linear system \eqref{eq:linear-sde} with  \oursanalysis matrix $\Ahat$ satisfies Eq.~\ref{eq:linear-sde-mean}.
By definition, 
$\cb_\pert = 0$ 
for the unperturbed drift $f$, 
so $\mub = \finff = - \widehat{\Ab}^{-1}\bb$
and thus
$\bb = - \Ahat \finff$.
Following the same reasoning, the perturbation vectors $\cb_q$ are given by
$\cb_\pert = - \bb -  \Ahat \finfpert$.

\medskip

\hfill $\blacksquare$

\section*{Proof of Theorem \ref{theorem:drift-transformed}}
\label{app:proof-theorem-drift-transformed}

To prove the statement, we first establish the following fact:

\begin{lemma}\label{lemma:speed-scaling}
Let $u: \RR^d \rightarrow \RR$ be a function such that $\exp(-2u(\xb))$ is integrable.
For any positive definite matrix $\scalingmat$,
the SDE
\begin{align}\label{eq:sde-speed-scaled}
	\dd\xb(t) = - \scalingmat \nabla u(\xb(t))\dd t +  \scalingmat^{1/2} \dd \WW(t)
\end{align}
has stationary distribution
\begin{align*}%
	p(\xb) \propto \exp(- 2 u(\xb))\, ,
\end{align*}
which is independent of the scaling matrix $\scalingmat$.
\end{lemma}

\paragraph*{Proof}
The Fokker-Planck equation \citep{sarkka2019applied} for the SDE \eqref{eq:sde-speed-scaled} states that the time evolution of the density $p(\xb, t)$ is given by
\begin{align*}
	\frac{\partial}{\partial t} p(\xb, t) = - \nabla \cdot J(\xb, t)\, ,
\end{align*}
where the probability flow is
\begin{align*}
	J(\xb, t) = - \scalingmat \nabla u(\xb)\, p(\xb, t) - \frac{1}{2} \scalingmat \nabla p(\xb, t)\, .
\end{align*}
A distribution $p(\xb)$ is stationary if and only if $\frac{\partial}{\partial t} p(\xb, t) = 0$, which by the Fokker-Planck equation is equivalent to $\nabla \cdot J(\xb) = 0$ for all $\xb$. 
A sufficient condition for this is that the probability current itself vanishes, i.e. $J(\xb) = \bzero$.
We verify that $p(\xb) \propto \exp(-2u(\xb))$ satisfies this condition.
For this choice of $p(\xb)$, we have
\begin{align*}
	\nabla p(\xb) = -2p(\xb)\nabla u(\xb)\, .
\end{align*}
Substituting into the expression for $J(\xb)$ yields
\begin{align*}
	J(\xb) 
	= - \scalingmat \nabla u(\xb)\, p(\xb) - \frac{1}{2} \scalingmat \nabla p(\xb) 
	= - \scalingmat \nabla u(\xb)\, p(\xb) + \scalingmat \nabla u(\xb)\, p(\xb) 
	= \bzero\, .
\end{align*}
Thus, $p(\xb) \propto \exp(-2u(\xb))$ is a stationary distribution, and this form is independent of $\scalingmat$.
The integrability condition ensures that the normalization constant is finite, so that $p(\xb)$ is a valid probability density.

\hfill $\blacksquare$

\bigskip

\noindent
We now prove the main result.
Given a transformation $\xb = \transform(\sbb)$,
the stochastic processes $\xb(t)$ is obtained from $\sbb(t)$ by
\begin{align*}
	\xb(t)
	= \transform(\sbb(t))
	= \tbias + \tmat \sbb(t)
	\, .
\end{align*}
By  It{\^o}'s Lemma \citep{oksendal2003stochastic}
applied to the general SDE form and $\transform$,
the SDE for $\xb(t)$ is 
\begin{align}
\begin{split}\label{eq:proof-transform-ito}
	\dd\xb(t) 
	&= \nabla \transform(\sbb(t))  \, \dd\sbb(t) \\
	&= \tmat\big(f(\sbb(t))\dd t + \sigma \dd\WW t\big)\\
	&= \tmat f(\sbb(t))\dd t + \sigma\tmat \dd\WW t\\
	&= \tmat f \big(\tmat^{-1}(\xb(t) - \tbias) \big)\dd t + \sigma\tmat \dd\WW t
\end{split}
\end{align}
This SDE induces the transformed stationary distribution $p(\xb)$.
However, the diffusion scale in Eq.~\ref{eq:proof-transform-ito} is $\sigma \tmat$, while we week the SDE with the same diffusion scale $\sigma$ as the original SDE.
Lemma \ref{lemma:speed-scaling} shows that a constant positive definite scaling $\scalingmat$ does not change the stationary distribution.
Hence, we apply the scaling factor $\scalingmat = \tmat^{-2}$ to Eq.~\ref{eq:proof-transform-ito}.
This yields
\begin{align*}
	\dd\widetilde{\xb}(t) 
	&= \tmat^{-1} f \big(\tmat^{-1}(\, \widetilde{\xb}(t) - \tbias \, ) \big)\dd t + \sigma \dd\WW t
	\, ,
\end{align*}
which induces the correct transformed stationary distribution with the original diffusion scale $\sigma$.

\hfill $\blacksquare$

\section*{Proof of Corollary \ref{corollary:clipr-vectors-transformed}}

By Theorem \ref{theorem:drift-transformed},
the drift $f^x $ is given by 
$f^x(\xb) = \tmat^{-1} f^s \big(\tmat^{-1}(\xb - \tbias)\big)$.
Since $\fzero_{f^x} = f^x(\bzero)$ by definition, we have
\begin{align*}
    \fzero_{f^x}
    = \tmat^{-1} f^s (-\tmat^{-1} \tbias) 
    \, .
\end{align*}
Moreover, the fixed point $\xb^*$ of $f^x$ satisfies
\begin{align*}
    f^x(\xb^*) 
    = \bzero
    \Leftrightarrow
    f^s \big(\tmat^{-1}(\xb^* - \tbias)\big)
    = \bzero 
\end{align*}
Since the fixed point $\sbb^*$ of $f^s$ satisfies $f^s(\sbb^*) = \bzero$, 
we have $\sbb^* = \tmat^{-1}(\xb^* - \tbias)$,
and conversely,
$\xb^* = \tbias + \tmat \sbb^*$.
Thus, $$\finf_{f^x} =  \tbias + \tmat\finf_{f^s} \, . $$

\hfill $\blacksquare$

\section*{Example: Linear SDEs and Gaussian distributions}

In this section, we illustrate Theorem \ref{theorem:clipr} and Corollary \ref{corollary:clipr-vectors} by applying them to linear SDEs
\eqref{eq:linear-sde},
for which the stationary distribution is known analytically.  
Consider an SDE system with drift $f^s(\sbb)$ that induces a Gaussian stationary distribution
$\sbb \sim \Ncal(\mub, \Pb^{-1})$.
We apply the elementwise transformation
\begin{align*}
	\xb = \transform(\sbb) = \tbias + \tmat \sbb \, ,
\end{align*}
where $\tmat$ is positive definite.
Under this transformation, $\xb$ remains Gaussian with
$\xb \sim \Ncal(\tbias + \tmat\mub, \tmat\Pb^{-1}\tmat)$.
Our goal is to derive the drift and causal effects of an SDE inducing the stationary distribution $p(\xb)$ using only the drift $f^s(\sbb)$ of an SDE inducing the stationary distribution $p(\sbb)$ and the transformation $\transform$, keeping the diffusion scale fixed.

We begin with a drift $f^s$ inducing the stationary distribution $p(\sbb)$.
For a Gaussian distribution, we have $\nabla \log p(\sbb) = - \Pb(\sbb - \mub)$.  
An SDE with stationary distribution $p(\sbb)$ is thus the Langevin diffusion with
\begin{align*}
	\dd \sbb(t) 
	= -\Pb(\sbb(t) - \mub) \dt + \sqrt{2} \dWt
	\, .
\end{align*}
This identifies the drift $f^s(\sbb)$ and the causal matrix $\Ab_s$ of the linear model 
\eqref{eq:linear-sde} 
as
\begin{align*}
	f^s(\sbb)
	= - \Pb(\sbb - \mub) 
	&&
	\Ab_s
	= - \Pb \, .	
\end{align*}
We now apply our results to obtain the parameters of an SDE inducing $p(\xb)$.

\bigskip

\noindent\textbf{\em Theorem \ref{theorem:drift-transformed}:}
Since $f^s(\sbb)$ is the gradient of a potential and the SDE admits a stationary distribution, 
Theorem \ref{theorem:drift-transformed} can be directly applied to obtain the drift $f^x(\xb)$ of an SDE inducing $p(\xb)$ as its stationary distribution with the same diffusion scale:
\begin{align*}
	f^x(\xb)
	&= \tmat^{-1} f^s \big ( \tmat^{-1} (\xb - \tbias )\big) 
	= - \tmat^{-1} \Pb \tmat^{-1} (\xb - \tbias - \tmat \mub) 
	\, .
\end{align*}
As expected, this matches the drift derived directly from the closed form of $p(\xb)$ by computing $\nabla \log p(\xb)$.
The corresponding matrix $\Ab_x$ of $f^x$ in Eq.~\ref{eq:linear-sde} is
\begin{align*}
	\Ab_x
	= -\tmat^{-1} \Pb\tmat^{-1} \, .
\end{align*}

\noindent\textbf{\em Corollary \ref{corollary:clipr-vectors-transformed}:}
Applying Corollary \ref{corollary:clipr-vectors-transformed}, the perturbation responses of $f^x$ are
\begin{align*}
	\fzero_{f^x}
    &= \tmat^{-1} f^s (-\tmat^{-1} \tbias)
    = \tmat^{-1} \Pb \tmat^{-1} ( \tbias + \tmat \mub)
    \\
    \finf_{f^x}
	&=  \tbias + \tmat\finf_{f^s}
	= \tbias + \tmat\mub
\end{align*}
As shown in Theorem \ref{theorem:clipr},
the causal matrix $\Ab_x$ of $f^x$ must satisfy
\begin{align*}
    \Ab_x\finf_{f^x} = - \fzero_{f^x}
    \, .
\end{align*}
Substituting the expressions for 
$\fzero_{f^x}$ and $\finf_{f^x}$,
we obtain the condition
\begin{align*}
    \Ab_x(\tbias + \tmat\mub) 
    &= - \tmat^{-1} \Pb \tmat^{-1} ( \tbias + \tmat \mub)
    \, ,
\end{align*}
confirming that the matrix $\Ab_x$ derived from $f^x(\xb)$ exactly satisfies the constraints specified by $\fzero_{f^x}$ and $\finf_{f^x}$ in Corollary \ref{corollary:clipr-vectors-transformed}.

\section*{Gene Interaction Classification}\label{app:gi-metrics}

We follow the framework by Norman et al.~\citep{norman2019exploring}
to classify gene pairs into gene interaction (GI) types.
Their approach measures GIs by comparing the mean expression of double gene perturbations to those of the constituent single gene perturbations.
Since the data by Wessels et al.~\citep{wessels2023efficient} does not contain sufficient single gene perturbations to cover a substantial number of double perturbations, our GI analysis focuses on the data by Norman et al.~\citep{norman2019exploring} 
(Table \ref{tab:datasets}).

Our experiments were performed on a selection of $d=\num{1000}$ highly variable genes (\methods).
Following Norman et al.,
we performed GI type classifications based on the $p=\num{570}$ genes with a minimum mean unique molecular identifier (UMI) count of \num{0.5}.
We define:
\begin{itemize}
	\item $\deltab_a \in \RR^p$: differential expression vector for perturbation $\pert_a$
	\item $\deltab_b \in \RR^p$: differential expression vector for perturbation $\pert_b$
	\item $\deltab_{a,b} \in \RR^p$: differential expression vector for the combinatorial perturbation of $\pert_a$ and $\pert_b$
\end{itemize}
Here, differential expression denotes
the difference in mean counts 
between the perturbation and control ($\pert_0$) data
after library standardization.
Let $c_a, c_b \in \RR$ be the regression coefficients obtained by fitting 
\begin{align*}
	[\deltab_{a,b}]_g = c_a [\deltab_{a}]_g + c_b [\deltab_{b}]_g + \epsilon_g \, ,
\end{align*}
for each gene $g \in \{1, \ldots, p\}$, where $\epsilon_g$ denotes the residual error.
The parameters are estimated using robust (Theil-Sen) regression to reduce sensitivity to outliers and skewness.
We use the implementation by SciPy \citep{scipy} with $10^5$ randomly selected data point pairs and $10^3$ iterations for computing the spatial median of the two coefficients.

Norman et al.\ define four metrics for characterizing GIs (Table~\ref{tab:si-gi-metrics}).
These metrics quantify: 
(i) the magnitude of $c_a$ and $c_b$; 
(ii) the similarity among $\deltab_{a}$, $\deltab_{b}$, and $\deltab_{a,b}$, measured via distance correlation~\citep{szekely2007measuring}; 
(iii) the quality of the linear fit for $\deltab_{a,b}$; 
and (iv) the equality of contribution of $\deltab_{a}$ and $\deltab_{b}$ in explaining $\deltab_{a,b}$. 
Based on these metrics, Norman et al.\ classify GIs into six categories: suppression, additivity, synergy, epistasis, redundancy, and neomorphism.

The original GI classifications by Norman et al.~\citep{norman2019exploring} were annotated based on GI metrics computed over the full gene set, whereas our analysis focused on $d=1000$ highly variable genes.
Although our metrics remain strongly correlated to those computed in the original study, preprocessing choices and gene selection affect their values and thus the GI classifications (Fig.~\ref{fig:comparison-gi-metrics}).
We therefore re-calibrated the classification thresholds for our gene set, following an approach similar to that of Roohani et al.~\citep{roohani2024predicting}.
Specifically, we determined the GI metric thresholds required to recover the original GI annotations given the metrics computed under our gene selection (for robustness considering 5--95 percentiles of the computed metrics) (Fig.~\ref{fig:comparison-gi-thresholds}).
The resulting thresholds are provided in Table~\ref{tab:si-gi-classifications} and were used to classify all available double gene perturbations.

\section*{Maximum Mean Discrepancy}\label{app:metrics}
We benchmarked methods by comparing 
a predicted set of $N$ samples 
$\widehat{\Yb}_\pert$
for perturbation $\pert$
to the $N_\pert$ observed post-perturbation samples 
$\Yb_{\pert}$,
following library size standardization and log transformation (\methods).
Our primary metric was the maximum mean discrepancy (MMD), which quantifies the discrepancy between two distributions \citep{gretton2012kernel}.
Specifically, we computed an unbiased empirical estimate of the MMD with kernel $k$ \citep{gretton2012kernel}:
\begin{align}
\begin{split}
\label{eq:mmd}
\mathrm{MMD}(\widehat{\Yb}_{\pert}, \Yb_{\pert})
=&
\frac{1}{N (N-1)}
\sum_{n=1}^N \sum_{m\neq n}^N  k(\widehat{\yb}^{(n)},\widehat{\yb}^{(m)})
+
\frac{1}{N_\pert (N_\pert-1)}
\sum_{n=1}^{N_\pert} \sum_{m\neq n}^{N_\pert}  k(\yb^{(n)},\yb^{(m)}) \\
&-
\frac{2}{N N_\pert}
\sum_{n=1}^{N} \sum_{m=1}^{N_\pert}  k(\widehat{\yb}^{(n)},\yb^{(m)})
\, .
\end{split}
\end{align}

\section*{Baselines}\label{app:baselines}

We evaluated \ourss against various baseline methods. 
For GEARS \citep{roohani2024predicting}\footnote{GEARS code repository: \url{https://github.com/snap-stanford/GEARS}}
and CPA \citep{lotfollahi2023predicting}\footnote{CPA code repository: \url{https://github.com/theislab/cpa}}, 
we used the original implementations by the authors. 
For SALT and PEPER \citep{gaudelet2024season}, we used our own implementations.
All baselines generated samples from the post-perturbation distribution by predicting a (mean) post-perturbation shift and then shifting samples from the control ($\pert_0$) data by the predicted offset.
SALT, PEPER, and GEARS are designed to predict means only, and for CPA, this was the recommended approach in the implementation.
As described in \methods, SALT predicts the post-perturbation shift of a double perturbation $\pert_a, \pert_b$ as the sum $\Delta_{\pert_a} + \Delta_{\pert_b}$ of its constituent single post-perturbation shifts $\Delta_{\pert_a}, \Delta_{\pert_b} \in \RR^d$. \citep{gaudelet2024season}.

PEPER extends SALT by training a neural network correction that is added to the SALT offset $\Delta_{\pert_a} + \Delta_{\pert_b}$.
Specifically, the method learns two multiplayer perceptrons (MLPs) $\phi_1$ and $\phi_2$ that predict the correction term as
$\phi_2(\phi_1(\Delta_{\pert_a}) + \phi_1(\Delta_{\pert_b}))$
using the central moment discrepancy loss \citep{zellinger2017central}.
The model was trained for \num{30000} steps with batch size $\num{256}$
using the Adam optimizer \citep{kingma2015adam}
with learning rate $0.001$ and weight decay tuned over 
$\{0.1, 0.003, 0.0001\}$. 
We also tuned the number of MLP hidden layers over $\{1, 2\}$ and hidden sizes over $\{256, 1024\}$.

CPA trains an autoencoder using a Gaussian reconstruction loss and an adversarial cross entropy loss that encourages a separation of latent state and perturbation information.
We tuned the latent state dimensionality over $\{64, 256\}$;
encoder, decoder, and adversarial classifier hidden size over $\{1024, 2048, 4096\}$,
weight decay over $\{10^{-5}, 10^{-8}\}$;
dropout probability of all networks over $\{0.0, 0.2\}$;
and the strength of the adversarial loss over $\{0.01, 1.0\}$. 
Training performed a maximum number of \num{1000} epochs, and all networks used \num{2} hidden layers.
The method has additional hyperparameters, for which we selected defaults based on recommendations in the original implementation by the authors.

GEARS trains a graph neural network (GNN) over gene and perturbation relationship graphs constructed from prior knowledge databases.
We optimized the GEARS model for \num{20} epochs
and tuned the batch size over $\{32, 256\}$,
the Pearson correlation threshold when constructing the co-expression graph over $\{0.4, 0.8\}$,
the maximum similar genes considered in the graphs over $\{5, 10, 20\}$,
and the number of GNN layers over $\{1, 2\}$.
The remaining hyperparameters were kept at defaults suggested by the original implementation.
Hyperparameter tuning is described in \methods.

\section*{Computational Resources}
We trained \ourss on an internal cluster
with \num{1} GPU (\num{20}GiB) and \num{4} CPUs (\num{24}GiB each for Wessels. et al. and Norman et al. datasets; \num{120}GiB each for Replogle et al. dataset).
Inferring the implicit prior densities was more computationally intensive 
($\sim$\num{9}h; inference step 1) 
than training the drift ($\sim$\num{1}h; inference step 2),
because of the large number of Monte Carlo samples $M$ needed to sufficiently reduce the bias of $\Lcal_\mathrm{IP}$ %
and achieve good fit of the priors.

\newpage
\begin{figure*}[p]
    \centering
    \includegraphics[
        width=\textwidth,
        trim={1pt 320pt 1pt 0pt}, clip
    ]{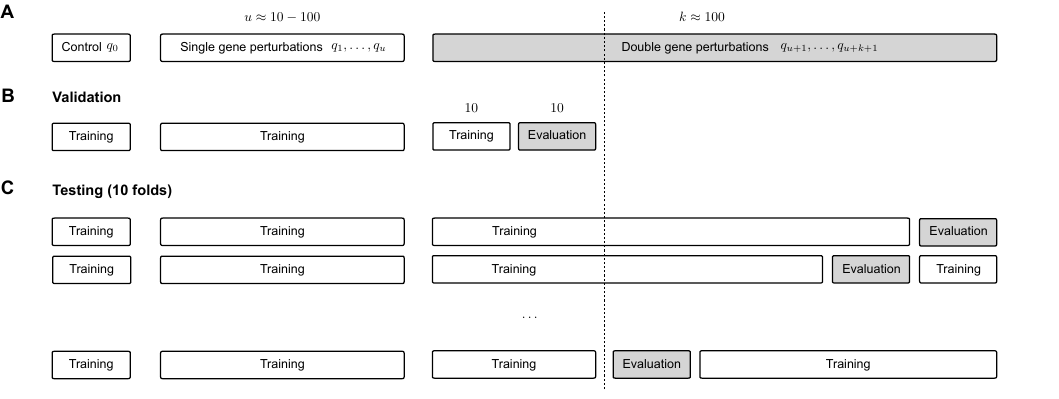}
    \caption{%
    \textbf{Training and evaluation splits of the combinatorial gene perturbation data by  Wessels et al.~\citep{wessels2023efficient} and Norman et al.~\citep{norman2019exploring} used for benchmarking \figref{fig:benchmarking}.}
    \subp{A}~Each dataset contains control samples ($\pert_0$), $u$ single gene perturbations, and $k$ double gene perturbations.
    Control and single gene perturbation samples are always in the training set.
    \subp{B} We reserved \num{20} double gene perturbations for hyperparameter tuning of each method (\methods), where \num{10} were included in the training set and \num{10} were used for evaluation.
    \subp{C} For benchmarking, we evaluated methods on \num{10} disjoint testing folds of the held-out double perturbations. 
    All perturbations used for hyperparameter tuning were included in the training set to avoid information leakage. 
    Combining all testing folds, methods generated held-out predictions for all $k - 20$ held-out double perturbations.
    }
    \label{fig:illustration-benchmark-splits}
\end{figure*}

\clearpage
\begin{figure*}[p]
    \centering
    \vspace{-20pt}
    \includegraphics[
        width=\textwidth,
        trim={0pt 0pt 0pt 0pt}, clip
    ]{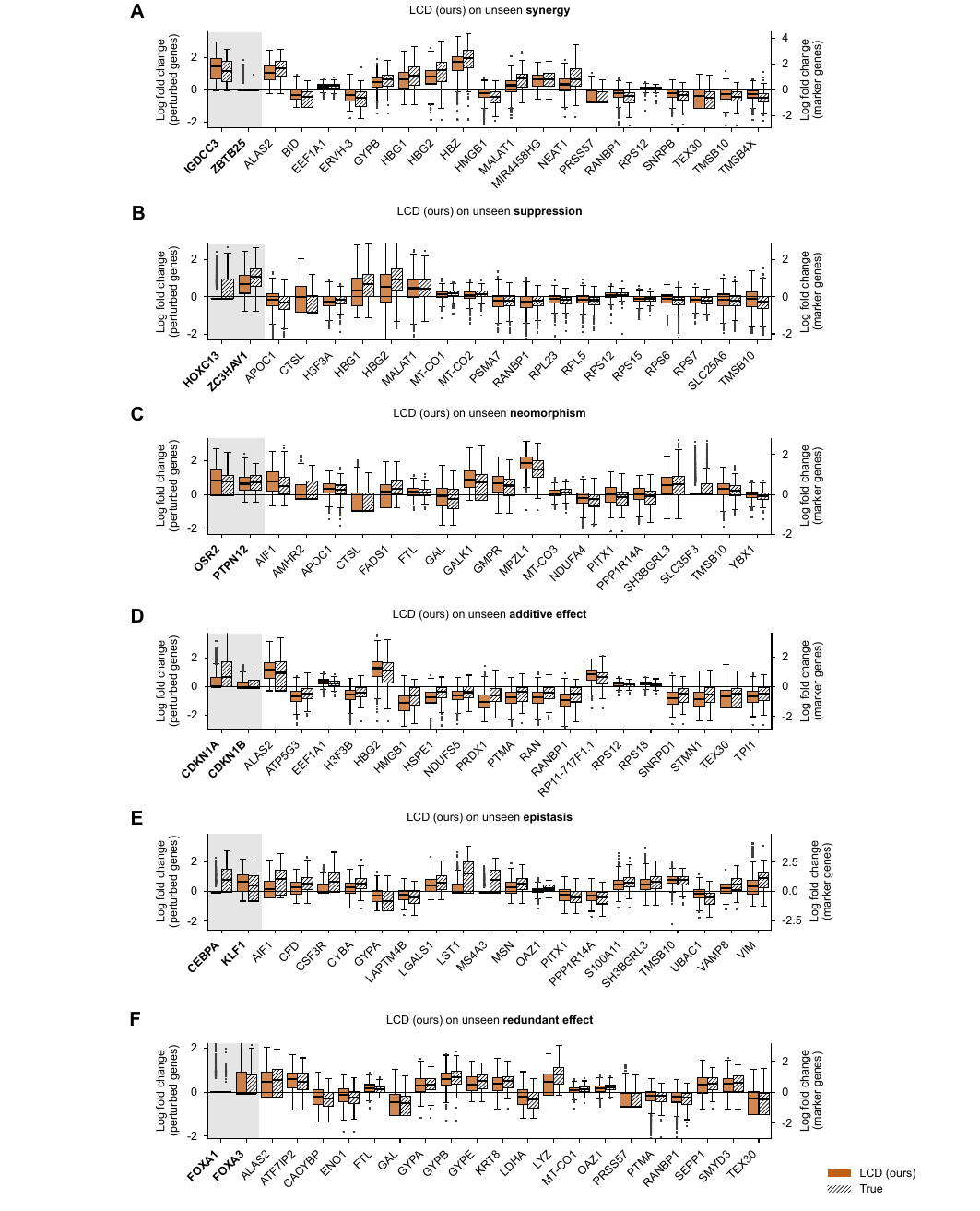}
    \caption{%
    \textbf{\ours predictions on two-gene perturbation combinations with different genetic interaction (GI) types.}
    Predicted vs.\ observed expression for top 20 \DE genes of a held-out perturbation with neomorphic  GI. 
    Grey section shows the perturbed genes (bold labels).
    All box plots show median and interquartile range.
    }
    \label{fig:de-gi-supplementary}
\end{figure*}

\clearpage
\begin{figure*}[p]
    \centering
    \includegraphics[
        width=\textwidth,
        trim={3pt 230pt 3pt 3pt}, clip
    ]{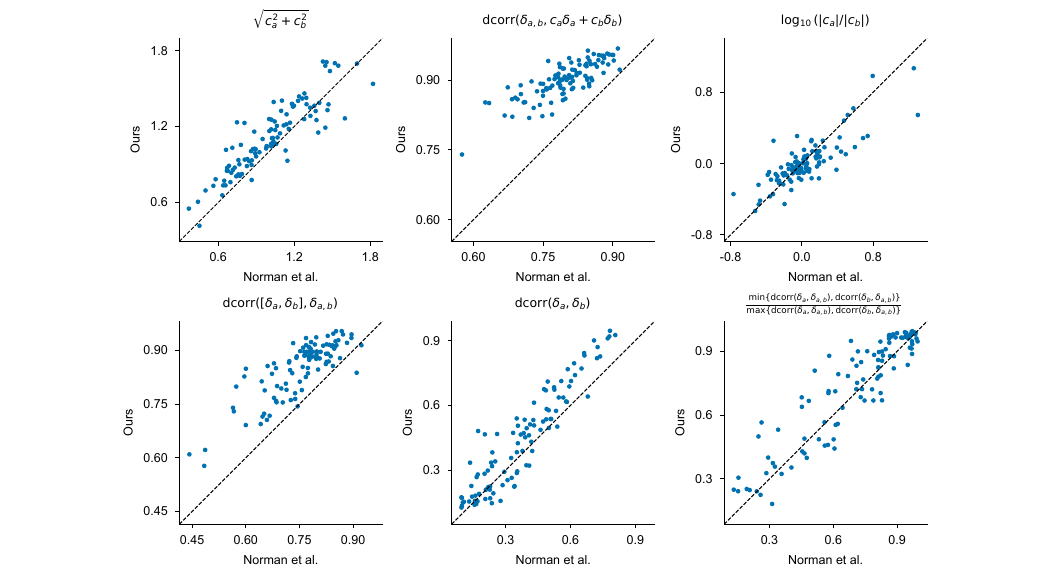}
    \caption{%
    \textbf{Comparison of gene interaction (GI) metrics between the original study \citep{norman2019exploring} and our experiments.}
    We selected $d=1000$ highly variable genes and applied specific preprocessing steps for quality control, which affect the computed GI metrics.
    Our results show strong correlation with those from the original study, motivating our decision to calibrate GI type thresholds based on the original annotations (\refappfig{fig:comparison-gi-thresholds}).
    Each point represents a gene pair for which a double perturbation and both constituent single perturbations are available, thus enabling GI metric computation.
    }
    \label{fig:comparison-gi-metrics}
\end{figure*}

\clearpage
\begin{figure*}[p]
    \centering
    \includegraphics[
        width=\textwidth,
        trim={3pt 245pt 3pt 3pt}, clip
    ]{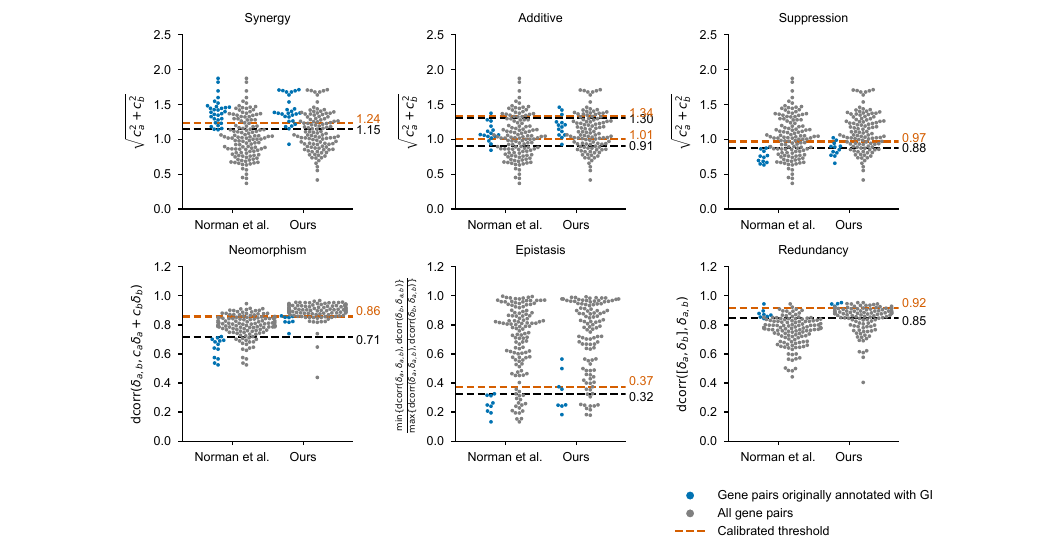}
    \caption{%
    \textbf{Calibration of gene interaction (GI) classification thresholds between the original study \citep{norman2019exploring} and our experiments.}
    Each panel corresponds to one of six GI categories, comparing the distribution of GI metrics from the original study (left) with those from our experiments (right).
    Blue points indicate gene pairs annotated with the respective GI type in the original study; grey points represent all available gene pairs.
	Black horizontal lines denote classification thresholds derived from the original annotations; red lines show thresholds calibrated to our metrics based on the scores of the originally annotated pairs.
	Thresholds are summarized in \refapptab{tab:si-gi-classifications}. 
    }
    \label{fig:comparison-gi-thresholds}
\end{figure*}

\clearpage
\begin{figure*}[p]
    \centering
    \includegraphics[
        width=\textwidth,
        trim={1pt 115pt 1pt 1pt}, clip
    ]{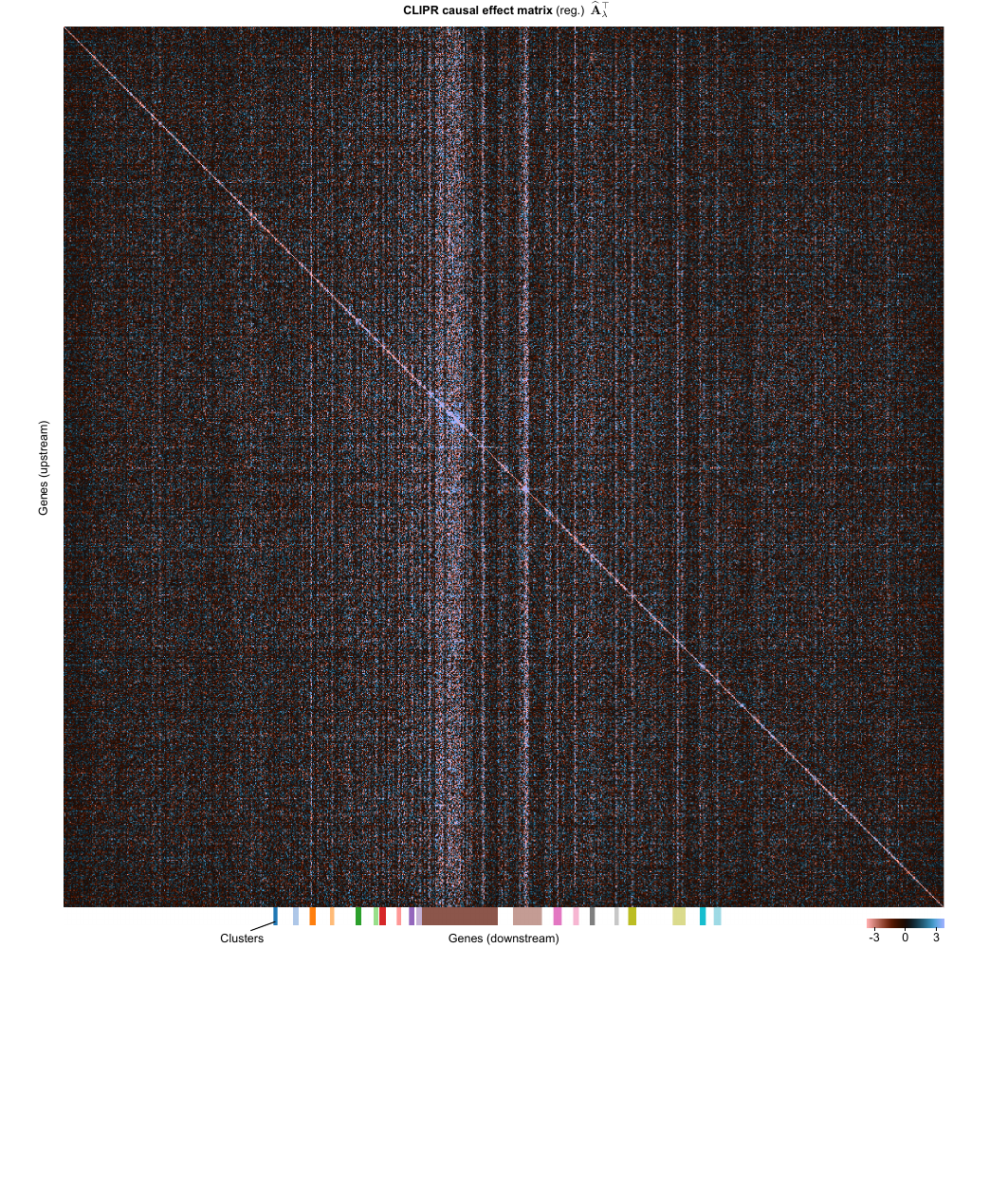}
    \caption{%
    \textbf{\ourscombo estimate of the direct gene-gene causal effects among all $d=1000$ modeled genes from the \perturbseq screen by Replogle et al.~\citep{replogle2022mapping}.}
    This plot complements \ffigref{fig:causality}[B], where only genes assigned to clusters are shown for visual clarity.
    Clusters indicate those analyzed in the enrichment analysis.
    The matrix is transposed into graph adjacency orientation, so that entry $(a,b)$ corresponds to the direct causal effect of gene $a$ on gene $b$.
    Tikhonov regularization: $\lambda = 0.01$ (see \methods).
    }
    \label{fig:full-resolution-causal-replogle}
\end{figure*}

\clearpage
\begin{figure*}[p]
    \centering
    \vspace{-25pt}
    \includegraphics[
        width=\textwidth,
        trim={1pt 10pt 1pt 3pt}, clip
    ]{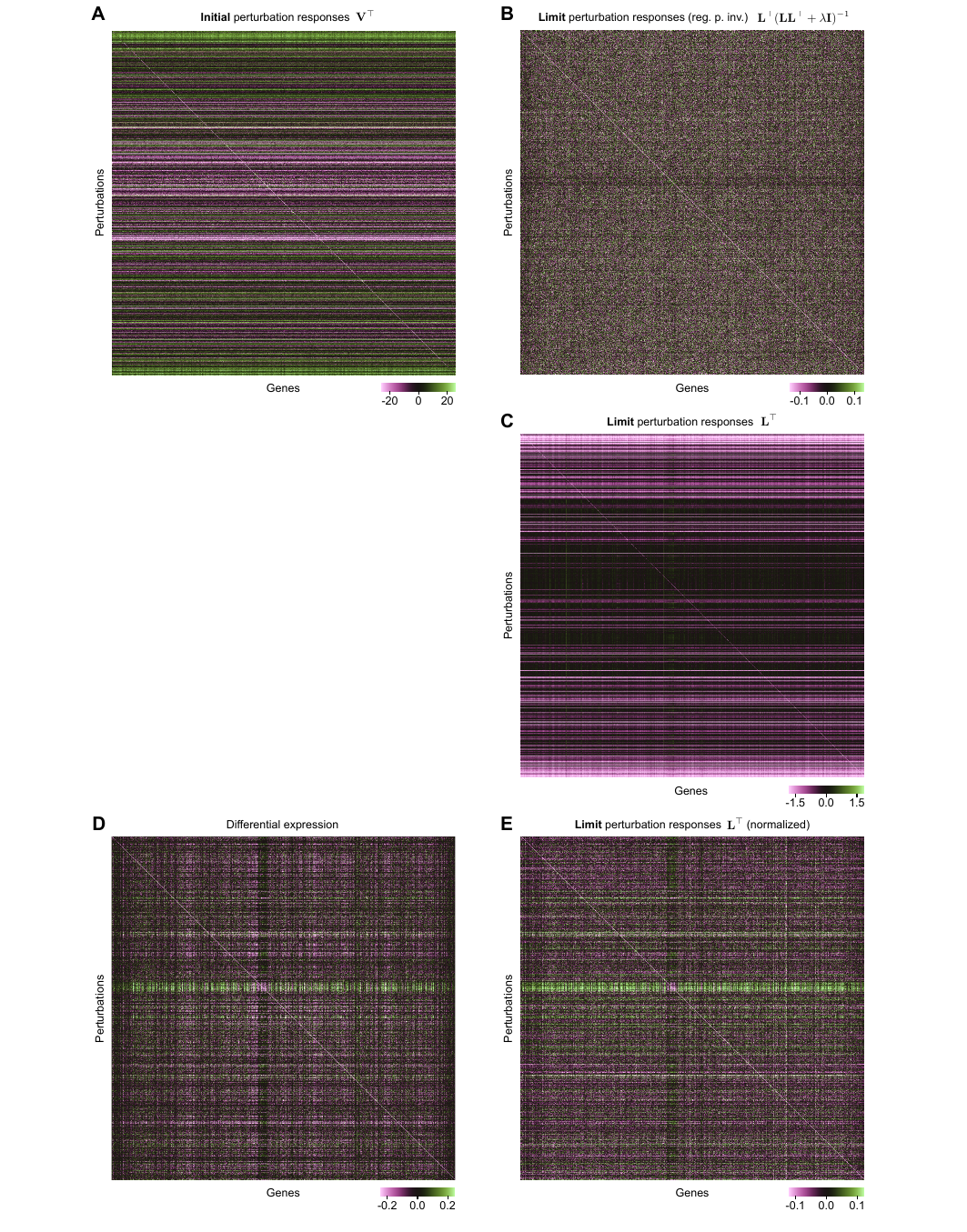}
    \caption{%
    \textbf{\oursanalysis components of the causal effects $\smash{\Ahat_\lambda}$ in \ffigref{fig:full-resolution-causal-replogle}.}
    This plot complements \ffigref{fig:causality}[C], where only genes assigned to clusters are shown.
    \subp{A}~Initial perturbation responses \eqref{eq:fzero-vector}.
    \subp{B} Tikhonov-regularized pseudoinverse of the limit perturbation responses. 
    The \oursanalysis matrix $\smash{\Ahat_\lambda}$ is equal to the (negative) product of (\textbf{A}, transposed) and \subp{B}.
    \subp{C} Limit perturbation responses \eqref{eq:finf-vector}.
    \subp{D} Differential expressions observed in the data.
    \subp{E} Limit perturbation responses, normalized 
    by passing values through the Poisson rate functions
    $\ratefunction_g$, %
    then applying the same library size standardization and log transformation used to preprocess the data.
    \subp{E} corresponds closely to \subp{D}.
    Matrices are transposed with perturbation indices on the first axis.
    Perturbation on row $g$ targets the gene on column $g$.
    Tikhonov regularization: $\lambda = 0.01$ (see \methods).
    }
    \label{fig:full-resolution-causal-replogle-mats}
\end{figure*}

\clearpage
\begin{table}[p]
\caption{
Perturbation statistics of all datasets after preprocessing and gene selection.
}\label{tab:datasets}
\centering
\begin{adjustbox}{max width=\linewidth}
\begin{threeparttable}
\centering
\begin{tabular}{lllcccccc}
Dataset
&
Perturbation 
&
\makecell{Cell\\line}
&
Cells
&
Genes
&
\makecell[l]{Median cells\\per perturb.}
&
\makecell{One-gene\\perturbs.}
&
\makecell{Two-gene\\perturbs.}
&
\makecell{Genes\\perturbed}
\\
\cmidrule{1-9}
Wessels et al.~\citep{wessels2023efficient}
& 
\makecell[l]{
Loss-of-function\\
(knockdown; CaRPool-seq)
}
&
THP-1
& \num{23722}
& \num{1000}
& \num{175}
& \num{5} 
& \num{104}  
& \num{25}
\\[2ex]
Norman et al.~\citep{norman2019exploring} 
& 
\makecell[l]{
Gain-of-function \\
(overexpression; CRISPRa)
}
&
K562
& \num{96804}
& \num{1000}
& \num{338}
& \num{98}
& \num{110}
& \num{98}
\\[2ex]
Replogle et al.~\citep{replogle2022mapping} 
& 
\makecell[l]{
Loss-of-function\\
(knockdown; CRISPRi)
}
&
K562
& \num{209462}
& \num{1000}
& \num{165}
& \num{1000}
& --
& \num{1000}
\\
\cmidrule{1-9}
\end{tabular}
\end{threeparttable}
\vspace*{0pt}
\end{adjustbox}
\end{table}

\clearpage
\begin{table}[p]
\caption{
Genetic interaction metrics following Norman et al.~\citep{norman2019exploring}.
}\label{tab:si-gi-metrics}
\centering
\begin{adjustbox}{max width=\linewidth}
\begin{threeparttable}
\centering
\begin{tabular}{lll}
GI metric 
& 
Symbol 
& 
Definition 
\\[0ex]
\cmidrule{1-3}
Magnitude 
& 
$\gimag$ 
& 
$\sqrt{c_a^2 + c_b^2}$ 
\\[2ex]
Similarity 
& 
$\gisim$ 
& 
$\dcor([\deltab_{a}, \deltab_{b}],\deltab_{a,b})$ 
\\[2ex]
Linear fit 
& 
$\gifit$ 
& 
$\dcor(\deltab_{a,b}, c_a \deltab_{a} + c_b \deltab_{b})$ 
\\[2ex]
Equality of contribution 
& 
$\giequality$ 
& 
$\displaystyle \frac{\min \{ \dcor(\deltab_{a},\deltab_{a,b}) ,  \dcor(\deltab_{b},\deltab_{a,b})\}}{\max \{  \dcor(\deltab_{a},\deltab_{a,b}),  \dcor(\deltab_{b},\deltab_{a,b})\}}$ 
\\[1ex]
\cmidrule{1-3}
\end{tabular}
\end{threeparttable}
\end{adjustbox}
\vspace*{0pt}
\end{table}

\clearpage
\begin{table}[p]
\caption{
Categories of genetic interactions 
based on the metrics in Table~\ref{tab:si-gi-metrics}
following Norman et al.~\citep{norman2019exploring}.
}\label{tab:si-gi-classifications}
\centering
\begin{adjustbox}{max width=\linewidth}
\begin{threeparttable}
\centering
\begin{tabular}{l@{\qquad}r@{\,}c@{\hspace{3pt}}l@{\hspace{3pt}}c@{\,}l@{\qquad}l}
Category 
& 
\multicolumn{5}{c}{Criterion}
& 
Description 
\\
\cmidrule{1-7}
Suppression 
& 
& & $\gimag$ & $<$ & $0.97$ 
& 
Magnitude of double is smaller than expected from individual singles 
\\[1ex]
Additivity
& 
$0.97$ & $<$ & $\gimag$ & $<$ & $1.34$ 
& 
Magnitude of double is approximately equal to combination of singles 
\\[1ex]
Synergy 
& 
$1.34$ & $<$ & $\gimag$ & & 
& 
Magnitude of double is larger than expected from individual singles 
\\[1ex]
Epistasis 
& 
& & $\giequality$ & $<$ & $0.37$ 
& 
One perturbation dominates the other in explaining the double profile 
\\[1ex]
Redundancy 
& 
$0.92$ & $<$ & $\gisim$ & & 
& 
Single profiles strongly explain the profile of the double 
\\[1ex]
Neomorphism 
& 
& & $\gifit$ & $<$ & $0.86$ 
& 
Linear model does not provide good fit of the double profile 
\\[0ex]
\cmidrule{1-7}
\end{tabular}
\end{threeparttable}
\vspace*{0pt}
\end{adjustbox}
\end{table}

\clearpage

\end{document}